\documentclass[12pt, letterpaper]{JHEP3}
\usepackage{amssymb, amsmath, amsopn, amsthm}
\usepackage{epsfig}
\usepackage{psfrag}
\usepackage{subfigure}

\newcommand{\eref}[1]{(\ref{#1})}

\newcommand{\fref}[1]{Figure~\ref{#1}}
\newcommand{\cref}[1]{Chapter~\ref{#1}}
\newcommand{\beq}{\begin{equation}}
\newcommand{\eeq}{\end{equation}}
\newcommand{\ba}{\begin{array}}
\newcommand{\ea}{\end{array}}
\newcommand{\bcenter}{\begin{center}}
\newcommand{\ecenter}{\end{center}}

\def\IB{\relax\hbox{$\inbar\kern-.3em{\rm B}$}}
\def\IC{\relax\hbox{$\inbar\kern-.3em{\rm C}$}}
\def\ID{\relax\hbox{$\inbar\kern-.3em{\rm D}$}}
\def\IE{\relax\hbox{$\inbar\kern-.3em{\rm E}$}}
\def\IF{\relax\hbox{$\inbar\kern-.3em{\rm F}$}}
\def\IG{\relax\hbox{$\inbar\kern-.3em{\rm G}$}}
\def\IGa{\relax\hbox{${\rm I}\kern-.18em\Gamma$}}
\def\IH{\relax{\rm I\kern-.18em H}}
\def\IK{\relax{\rm I\kern-.18em K}}
\def\IL{\relax{\rm I\kern-.18em L}}
\def\IP{\relax{\rm I\kern-.18em P}}
\def\IR{\relax{\rm I\kern-.18em R}}
\def\IZ{\relax\ifmmode\mathchoice
{\hbox{\cmss Z\kern-.4em Z}}{\hbox{\cmss Z\kern-.4em Z}}
{\lower.9pt\hbox{\cmsss Z\kern-.4em Z}}
{\lower1.2pt\hbox{\cmsss Z\kern-.4em Z}}\else{\cmss Z\kern-.4em Z}\fi}
\def\II{\relax{\rm I\kern-.18em I}}

\def\sCC{{\kern 0.27em\vrule height1.45ex width0.03em depth0em
          \kern-0.30em\rm C}}
\def\C{{\mathchoice
  {\sCC}
  {\sCC}
  {\kern 0.225em \vrule height1.05ex width0.025em depth0em \kern-0.25em \rm C}
  {\kern 0.180em \vrule height0.78ex width0.02em depth0em \kern-0.2em \rm C}
        }}
\def\sHH{{\rm I\kern-.16em{}H}}
\def\H{{\mathchoice
  {\sHH}
  {\sHH}
  {\rm I\kern-.13em{}H}
  {\rm I\kern-.13em{}H} }}
\def\sNN{{\rm I\kern-.16em{}N}}
\def\N{{\mathchoice
  {\sNN}
  {\sNN}
  {\rm I\kern-.12em{}N}
  {\rm I\kern-.10em{}N} }}
\def\sPP{{\rm I\kern-.16em{}P}}
\def\P{{\mathchoice
  {\sPP}
  {\sPP}
  {\rm I\kern-.12em{}P}
  {\rm I\kern-.10em{}P} }}
\def\sQQ{{\kern 0.27em \vrule height1.45ex width0.03em depth0em
          \kern-0.30em \rm Q}}
\def\Q{{\mathchoice
        {\sQQ}
        {\sQQ}
  {\kern 0.225em \vrule height1.05ex width0.025em depth0em \kern-0.25em \rm Q}
  {\kern 0.180em \vrule height0.78ex width0.020em depth0em \kern-0.20em \rm Q}
        }}
\def\sRR{{\rm I\kern-0.16em{}R}}
\def\R{{\mathchoice
  {\sRR}
  {\sRR}
  {\rm I\kern-0.12em{}R}
  {\rm I\kern-0.10em{}R} }}
\def\sZZ{{\rm Z\kern-0.32em{}Z}}
\def\Z{{\mathchoice
  {\sZZ}
  {\sZZ} 
  {\rm Z\kern-0.3em{}Z}     
  {\rm Z\kern-0.25em{}Z} }}  
\def\ZZZ{{\rm Z\kern-0.24em{}Z}}
\def\sII{{\rm I\kern-0.16em{}I}}
\def\I{{\mathchoice
  {\sII}
  {\sII}
  {\rm I\kern-0.12em{}I}
  {\rm I\kern-0.10em{}I} }}

\def\Tr{{\rm Tr}}

\def\inbar{\,\vrule height1.5ex width.4pt depth0pt}
\font\cmss=cmss10 \font\cmsss=cmss10 at 7pt

\def\smiley{\hbox{\large$\bigcirc$\hspace{-0.80em}\raise.2ex
\hbox{$\cdot\cdot$}\kern-.61em\lower.2ex\hbox{\scriptsize$\smile$}}\ }
\def\frowny{\hbox{\large$\bigcirc$\hspace{-0.80em}\raise.2ex
\hbox{$\cdot\cdot$}\kern-.635em\lower.2ex\hbox{\scriptsize$\frown$}}\ }

\def\I{{\rlap{1} \hskip 1.6pt \hbox{1}}}

\makeatletter
\let\hangafter\@hangfrom
\makeatother

\newcommand{\drawsquare}[2]{\hbox{%
\rule{#2pt}{#1pt}\hskip-#2pt
\rule{#1pt}{#2pt}\hskip-#1pt
\rule[#1pt]{#1pt}{#2pt}}\rule[#1pt]{#2pt}{#2pt}\hskip-#2pt
\rule{#2pt}{#1pt}}

\newcommand{\fund}{\raisebox{-.5pt}{\drawsquare{6.5}{0.4}}}
\newcommand{\antifund}{\overline{\fund}}

\title{SUSY breaking mediation by D-brane instantons}

\author{Matthew Buican and Sebasti\'an Franco

\\
~\\
Joseph Henry Laboratories, Princeton University \\
Princeton, NJ 08544, USA \\ \vspace{0.3cm}

\email{mbuican, sfranco@Princeton.EDU}\\
}

\abstract{It is well known that D-brane instantons can generate contributions to the effective superpotential of gauge theories living on D-branes which are perturbatively forbidden by global $U(1)$ symmetries. We extend this idea to theories with supersymmetry breaking, studying the effect of D-brane instantons stretched between the SUSY-breaking and visible sectors. Analogously to what happens in the SUSY case, this mechanism can give rise to perturbatively forbidden soft terms (among other effects). We introduce and discuss general properties of instanton mediation. We illustrate our ideas in simple Type IIB toroidal orientifolds. As a bi-product, we present a string theory realization of a Polonyi hidden sector.}

\preprint{PUPT-2272}

\def\be{\begin{equation}}
\def\ee{\end{equation}}
\def\bea{\begin{eqnarray}}
\def\eea{\end{eqnarray}}

\newcommand{\id}{\bf 1}
\newcommand{\diag}{{\rm diag}}

\newcommand{\symm}{~\raisebox{-.5pt}{\drawsquare{6.5}{0.4}}\hskip-0.4pt%
        \raisebox{-.5pt}{\drawsquare{6.5}{0.4}}~}
        
\newcommand{\antiasymm}{~\overline{\raisebox{-3.5pt}{\drawsquare{6.5}{0.4}}\hskip-6.9pt%
        \raisebox{3pt}{\drawsquare{6.5}{0.4}}}~}

\newcommand{\antisymm}{~\overline{\raisebox{-.5pt}{\drawsquare{6.5}{0.4}}\hskip-0.4pt%
        \raisebox{-.5pt}{\drawsquare{6.5}{0.4}}}~}

\begin{document}
\tableofcontents


\section{Introduction}

Supersymmetry (SUSY) is one of the most concrete ideas for stabilizing the large hierarchy between the weak scale ($\sim$100 GeV) and the Planck scale ($\sim10^{19}$ GeV). However, SUSY is not a symmetry of the low energy world we inhabit or even of the world that has been explored by high energy experiment thus far. As a result, we require that supersymmetry be broken at scales above those already probed. In order not to spoil the supersymmetric stabilization of the large hierarchy between weak and gravitational physics, however, we demand that the scale of supersymmetry breaking be roughly $\mathcal{O}$(1-10 TeV).

Furthermore, since SUSY requires the existence of a \lq superpartner' of opposite statistics and equal mass for each field in the Standard Model, the corresponding masses for these superpartners must be large enough so that they could have avoided detection in experiment thus far. On general grounds, one can then show that a model of SUSY breaking in which the Supersymmetric Standard Model fields themselves spontaneously break SUSY is ruled out since otherwise some of the scalar quarks (\lq squarks' for short) would have a mass smaller than the up quark mass \cite{Dimopoulos:1981zb}.

Thus, in order to have phenomenologically viable models, we are left with a situation in which SUSY must be explicitly broken in the interactions of the Standard Model fields. One can then adopt a paradigm in which the theories of interest consist of two sectors: one, a so-called \lq visible' sector that corresponds to the MSSM or an extension of the MSSM and a separate \lq hidden' sector where (spontaneous) SUSY breaking occurs. In order to have a complete model, one must specify a mechanism that communicates, or, in the language of the literature, \lq mediates' the SUSY breaking of the hidden sector to the visible sector. Known examples of such mediation mechanisms include gauge mediation, where SUSY breaking is communicated via loops of fields charged under the visible sector gauge group that acquire a non-supersymmetric mass due to couplings with fields that acquire SUSY breaking vevs and gravity mediation, where one generates couplings between the hidden and visible sector from integrating out Planck scale states.

Now, any mechanism that breaks SUSY must not reintroduce quadratic divergences into the effective action of a supersymmetric extension of the Standard Model through renormalizable operators.  SUSY breaking that satisfies this condition is termed \lq soft' SUSY breaking and can be parametrized by a set of so-called \lq soft' terms in the Lagrangian \cite{Girardello:1981wz}.\footnote{The mediation mechanisms discussed in the previous paragraph are examples of mechanisms that generate soft SUSY breaking.} In general, the number of such terms is large. Indeed, even in the relatively simple case of the Minimal Supersymmetric Standard Model (MSSM), the number of soft terms is 105. This plethora of possible new parameters in the Lagrangian seems to introduce unwanted arbitrariness. Nevertheless, most of the possible choices of soft terms give rise to unacceptable flavor and CP violation, so we are actually left with a constrained slice of parameter space. A major goal of model building is to come up with SUSY breaking and mediation mechanisms that result in such constrained soft terms.

In addition to successfully satisfying the various phenomenological requirements listed above, an appealing mediation mechanism should have distinctive soft signatures that can be picked out in collider experiments, and, ideally, a simple characterization of its parameter space.\footnote{Recent work on gauge mediation shows that, under a careful but rather broad definition, it satisfies these principles, since it generically features special linear relations between scalar masses and also has a tightly-controlled parameter space governed by a small number of hidden sector current-current correlators \cite{GGM}.} One might then even hope that the soft signature of the mediation mechanism could shed some light on the physics of the SUSY breaking sector.

String theory compactifications provide a natural and consistent laboratory in which to better understand the physics of theories with visible and hidden sectors. In these constructions, the two sectors correspond to different sets of (anti) D-branes separated in the extra dimensions. Various mediation mechanisms can then communicate SUSY breaking. They can be classified according to the string sector involved. We can have, for example: open string mediation (gauge mediation), closed string mediation (gravity mediation) and open/closed mixed mediation (RR $p$-form topological mediation, which uses RR $p$-forms to couple $U(1)$ gauge fields in the visible and hidden sectors \cite{Verlinde:2007qk}). More importantly, string theory furnishes a geometrical interpretation of the various SUSY breaking parameters and hence could lead to additional insight into the SUSY breaking physics that is difficult to obtain from field theoretic techniques alone.

As the above examples demonstrate, string theory comes with a whole host of unique objects that can be used in constructing mediation mechanisms. The goal of this paper is to describe another such mechanism. In particular, we will find that our mechanism has some rather unique properties that do not follow from the standard phenomenological literature on SUSY breaking mediation.

In particular, we will study the effects of Euclidean D-branes localized at a point in the non-compact four dimensions---so-called \lq D-brane instantons'--- stretching between the hidden and visible sectors.\footnote{In the last year and a half, there has been a surge in the study of D-brane instantons, mainly due to their ability to generate superpotential couplings that are perturbatively forbidden by $U(1)$ symmetries. Many applications
have been investigated, such as the cure of runaway directions in models that dynamically break SUSY \cite{Florea:2006si}, 
neutrino masses and mu terms \cite{Blumenhagen:2006xt,Ibanez:2006da,Ibanez:2007rs,Buican:2006sn,Cvetic:2007ku}, R-symmetry breaking and metastability \cite{Buican:2007is}, Yukawa couplings in GUT models \cite{Blumenhagen:2007zk} and SUSY breaking models with and without 
non-abelian gauge dynamics \cite{Argurio:2007qk,Aharony:2007db}. In this paper, we take the natural step of extending these ideas to the non-supersymmetric realm.} Upon integrating over the massless, charged zero mode strings stretching between the D-instanton and the hidden and visible sectors, we generate operators that couple the two sectors.\footnote{It is important to notice that the strength of D-brane instantons is not related to the strength of any MSSM instanton. Hence, they can be much less suppressed.} Roughly speaking, if the hidden sector breaks SUSY, these operators can then generate soft terms for the visible sector fields. In this sense, the D-brane instantons mediate SUSY breaking. Furthermore, our mediation mechanism has no known field theoretical analog.
More generally, we expect D-brane instanton mediation to be present in a variety of string theory constructions. Hence, it deserves to be studied, regardless of whether it is the dominant mediation mechanism or not.

In this paper, we find the following simple characterization of D-instanton mediation:
\begin{itemize}
\item D-instantons can generate soft terms that are perturbatively forbidden by $U(1)$ symmetries and hence cannot be generated by gauge mediation. They can also produce couplings that do not violate any $U(1)$ global symmetry, as in the example in section 4.2 and general models discussed in section 6.

\item There is a natural hierarchy of soft terms that is parameterized by the volumes wrapped by the corresponding D-instantons.

\item One can easily write down examples where the geometric conditions for SUSY breaking imply a particular hierarchy of soft terms from the mediating instantons.

\item D-instanton mediation is sensitive to the details of the SUSY breaking sector. In particular, if the SUSY breaking sector is a D-brane gauge theory that dynamically breaks SUSY, then D-brane instantons generate chiral gauge invariant operators that correspond to their orientation (or, more precisely, the homology cycle they wrap). In particular, some SUSY breaking hidden sectors fail to generate soft terms via this form of mediation since the mediating D-instantons project onto a subspace of vanishing chiral gauge invariants.
\end{itemize}

From a model building perspective, the last three points are potentially quite interesting. Significantly, they follow rather simply from the fact that our mediators have a clear geometrical interpretation. 

Let us briefly summarize the plan of this paper. In the next section, we will discuss in much greater detail the idea behind D-instanton mediation. Then we will quickly introduce the machinery of toroidal orientifolds with D-branes and D-instantons as a warm-up for some specific examples we then engineer in this setup. We will conclude with a brief and admittedly incomplete phenomenological discussion.

One of the main issues we do not address in this paper, but should certainly be studied, is the stabilization of compactification moduli, the dilaton and D-brane moduli after SUSY breaking. We have chosen to focus on the question of whether SUSY breaking
mediating interactions can be generated by D-brane instantons, under the assumption that it can be disentangled from
moduli stabilization. With this as our main goal, we do not try to engineer fully realistic or complete models of D-instanton mediation, but rather we discuss the basics of this mechanism in a few simple and illustrative toy examples. We leave a more detailed analysis to future work.

\section{General idea}

\subsection{Coupling visible and hidden sectors via D-brane instantons}

First, we will review the basics of how to generate chiral operators from Euclidean Dp-branes (Ep-branes for short). Since some of the soft terms (e.g., A-terms, B-terms, etc.) will be generated by products of chiral operators, we will be interested in this well-studied case. However, we will also discuss the possibility of generating non-chiral operators from instanton anti-instanton combinations (or, simply from single, non-BPS instantons \cite{Cvetic:2008ws,GarciaEtxebarria:2008pi}) as well, since such operators can give rise to non-holomorphic soft masses.

\begin{figure}
\begin{center}
\psfrag{X}[cc][][1]{$X_{ij}$}
\psfrag{SUi}[cc][][.9]{$SU(N)_i$}
\psfrag{SUj}[cc][][.9]{$SU(N)_j$}
\psfrag{a}[cc][][1]{$\alpha$}
\psfrag{b}[cc][][1]{$\beta$}
\includegraphics[width=4cm]{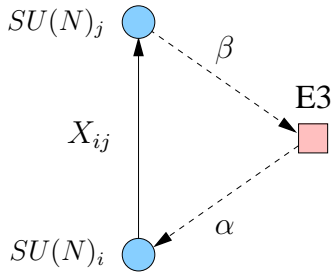}

\caption{Extended quiver diagram for the basic E-brane configuration that generates a superpotential contribution. Dotted arrows indicate charged fermionic zero modes. The figure presents the simplest case, in which the fermionic zero modes couple to a single bifundamental field between a pair of nodes. In the generic situation, charged zero modes can couple to more general operators, associated with an open path in the quiver.}
\label{quiver_instanton}
\end{center}
\end{figure}

We begin by imagining a space-filling D-brane sector that the various Ep-branes interact with. This could be a set of fractional branes at a singularity or a set of intersecting D-brane stacks. The physics of the Ep-brane interactions with the D-brane sector can be encoded in an extended quiver diagram of the form shown in \fref{quiver_instanton}.

The circles denote two of the gauge groups that are part of a larger quiver living on the set of space-filling D-branes. $X_{ij}$ corresponds to a combination of chiral fields transforming in the bifundamental representation of $SU(N)^{(i)} \times SU(N)^{(j)}$. The ranks of both gauge groups must be the same in order to have a non-vanishing instanton contribution. Note that $X_{ij}$ can simply be a single bifundamental field or, more generally, a product of the form $X_{ij}=X_{i\, k_1}X_{k_1 k_2} \ldots X_{k_n j}$, where we sum over the intermediate color indices. In section 6, we discuss this possibility in more detail. The square node in the extended quiver indicates the Ep-brane. Charged fermionic zero modes, represented by the Grassman variables $\alpha$ and $\beta$, arise at the intersections between the spacefilling D-branes and the instanton. The instanton action then contains a term of the form

\beq
L=\alpha_{i} X_{ij}\beta_{j} ~,
\label{cubic_coupling_instanton}
\eeq
Let us now discuss the neutral fermionic zero modes---these arise from strings that have both ends on the instanton. We will focus on orientifolded Calabi-Yau compactifications with D-branes, leading to ${\mathcal N}=1$ SUSY in 4d. Since the Ep-brane breaks 1/2 of the SUSY it therefore has two fermionic zero modes, the goldstinos, living on it---these are represented by the Grassman variables, $\theta^{\alpha}$. Generically, there are two additional fermionic zero modes on the instanton due to the `accidental' ${\mathcal N}=2$ SUSY seen by the Ep-Ep sector. This issue was first discussed and clarified through explicit computations in some orbifold models in \cite{Argurio:2007vqa,Bianchi:2007wy}. In order to saturate the superspace measure and generate a non-vanishing contribution to the superpotential, there must be only two neutral fermionic zero modes on the instanton. A straightforward way of getting rid of the accidental neutral zero modes is to project them out by placing the instanton on top of an orientifold plane with an $O(1)$ projection.\footnote{We can also consider using 3-form fluxes to lift the additional fermionic zero modes --- see \cite{Blumenhagen:2007bn} and \cite{Bergshoeff:2005yp} for a further discussion of this point. We will later comment on this scenario. Another possibility would be to consider a two-instanton contribution where the extra neutral zero modes are lifted by interactions between the two instantons \cite{GarciaEtxebarria:2007zv,Cvetic:2008ws,GarciaEtxebarria:2008pi}.} Then, after a straightforward Grassman integration over the charged zero modes, we obtain the following contribution to the 4d effective superpotential

\beq
W_{inst}=M_s^{3-N}e^{-V_{\Sigma}/g_s}\det X_{ij} ~.
\label{W_inst}
\eeq
where $V_{\Sigma}$ is the volume in string units of the cycle, $\Sigma$, wrapped by the instanton. In this expression and future ones, we omit a numerical multiplicative constant that we assume to be of ${\mathcal O}(1)$.

In orientifold singularities, the $SU(N)^{(i)}$ and $SU(N)^{(j)}$ nodes might be identified by the orientifold. In this case, depending on the charge of the corresponding O-plane, $X_{ij}$ is projected into a 2-index (conjugate) symmetric or antisymmetric representation of the resulting single $SU(N)$ factor. Furthermore, we get a single (anti)fundamental fermionic zero mode $\alpha$. In this case, the instanton action contains the coupling

\beq
L=\alpha_a X^{ab} \alpha_b ~.
\label{cubic_coupling_instanton_orientifold}
\eeq
where $X$ transforms in the $\antisymm$ or $\antiasymm$ representations, while $\alpha$ transforms in the $\fund$ representation (of course, we could also have the conjugate representations). We write the color indices $a$ and $b$ explicitly---they should not be confused with the quiver node labels $i$ and $j$. Integrating over $\alpha$ we get a contribution to the effective potential that takes the form

\beq
W_{inst}=M_s^{3-N/2}e^{-V_{\Sigma}/g_s}\sqrt{\det X} ~.
\label{W_inst_orientifold}
\eeq

Now, the gauge groups of quiver theories that arise on D-branes are actually $U(N)=SU(N)\times U(1)$. Any operator that does not correspond to a closed oriented path in the quiver is charged under some of these $U(1)$ symmetries. Thus, we conclude that
the instanton generated couplings \eref{W_inst} and \eref{W_inst_orientifold} are perturbatively forbidden by these $U(1)$ symmetries. 

Finally, let us also note that it may be possible to generate instanton-induced corrections to the holomorphic gauge kinetic functions of the various nodes in the quiver \cite{Akerblom:2007uc,Blumenhagen:2008ji}
\beq\label{Gkincorr}
e^{-V_{\Sigma}/g_s}W_{\alpha}W^{\alpha}
\eeq
Such corrections can arise from BPS Ep-branes that have additional neutral fermionic (non-Goldstino) zero modes and no massless modes charged under the corresponding $SU(N)$ factor (or, indeed, under any of the other gauge groups present). In the specific examples considered in \cite{Akerblom:2007uc} and \cite{Blumenhagen:2008ji} such corrections arise in the world-volume theory of D6-branes that interact with E2-branes wrapping cycles with a 1-dimensional 1-homology, i.e., $b_1(\Sigma)=1$. These E2-branes then have the requisite additional pair of neutral fermionic zero modes (after orientifolding) that allows them to generate corrections to the D6-gauge kinetic functions.

\begin{figure}
\begin{center}
\includegraphics[width=4.5cm]{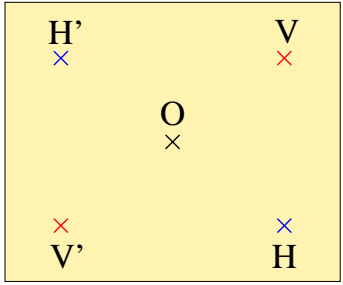}
\caption{The basic configuration for mediation. It consists of visible and hidden sectors $V$ and $H$, their orientifold images $V'$ and 
$H'$, and and O-plane $O$. All of them are intersected by a Euclidean D-brane, depicted in yellow.}
\label{basic_config}
\end{center}
\end{figure}

An interesting spectrum of new possibilities arises when we generalize the class of configurations we have just discussed to ones where an instanton can intersect multiple sets of D-branes that are spatially separated. For concreteness, we will specialize our discussion to compact type IIB orientifolds, in which we will study the effect of E3-branes wrapped over compact 4-cycles, $\Sigma$. \fref{basic_config} shows the simplest situation in which an E3-brane intersects only an O-plane and two D-brane sectors (plus their images). Motivated by our goal of investigating possible SUSY mediating effects in this class of setups, let us denote the two sectors visible and hidden. The hidden sector might involve anti D-branes, although we will not explore this possibility. It is possible for the visible or hidden sector and its image to collapse on top of the O-plane. Our previous discussion, regarding fermionic zero modes on the instanton and their projection via an orientifold, applies to this case without changes. 

The configuration can be more complicated than the simplest one, with the E-brane intersecting additional sectors with (anti) D-branes. These extra intersections result in additional insertions of 4d fields. 
This is an interesting direction that is worth studying. From now on, however, we restrict our discussion to some clean models in which this situation does not arise. 

\begin{figure}
\begin{center}
\psfrag{XV}[cc][][.9]{$X_{ij}^{(V)}$}
\psfrag{XH}[cc][][.9]{$X_{kl}^{(H)}$}
\psfrag{SUi}[cc][][.9]{$SU(N)_i$}
\psfrag{SUj}[cc][][.9]{$SU(N)_j$}
\psfrag{SUk}[cc][][.9]{$SU(N')_k$}
\psfrag{SUl}[cc][][.9]{$SU(N')_l$}
\includegraphics[width=7cm]{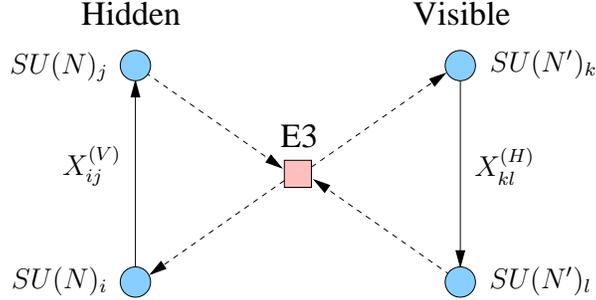}

\caption{Extended quiver diagram for the basic mediating configuration. Dotted arrows indicate charged fermionic zero modes. The figure presents the simplest case in which these fermionic zero modes couple to single bifundamental fields between pairs of nodes in the visible and hiden sectors. Generically, charged zero modes can couple to more general operators, associated with open paths in the quiver.}
\label{quiver_general}
\end{center}
\end{figure}

Once again, the configuration can be captured by an extended quiver as shown in \fref{quiver_general}. 
In this case we have two pairs of nodes, representing pairs of gauge groups in the visible and hidden sectors. In addition, we have fermionic zero modes $\alpha_V$, $\beta_V$, $\alpha_H$ and $\beta_H$ connecting the instanton to bifundamental operators $X^{(V)}_{ij}$ and $X^{(H)}_{kl}$ via the couplings

\beq\label{EpDp}
L=\alpha_V X^{(V)}_{ij} \beta_V + \alpha_H X^{(H)}_{kl} \beta_H ~.
\eeq
Upon integrating over the E3 zero modes, we generate contributions to the superpotential of the form
\be\label{WI}
W_{H/V}=M_s^{3-d_H-d_V}e^{-V_{\Sigma}/g_s} \, \mathcal{O}_H\mathcal{O}_V
\ee
where

\beq
\mathcal{O}_V= \det X^{(V)}_{ij},  \ \ \ \ \ \ \ \mathcal{O}_H= \det X^{(H)}_{kl} ~.
\eeq
In the previous discussion, we have implicitly assumed that gauge groups are not identified by orientifold projections. The obvious modifications along the lines of \eref{cubic_coupling_instanton_orientifold} and \eref{W_inst_orientifold} apply if the hidden and/or the visible sectors involve orientifold identifications.

Using similar reasoning, let us also note that by taking instantons that generate the corrections, written in (\ref{Gkincorr}), to the gauge kinetic functions of the space-filling D-branes and allowing them to intersect the hidden sector, we can generate the following operators involving the visible sector field strength superfields
\beq\label{Gkincorr2}
W_{H/V}=e^{-V_{\Sigma}/g_s}\mathcal{O}_HW_{V\alpha}W^{\alpha}_V
\eeq

Now that we have discussed the generation of chiral operators from Euclidean D-branes stretching between the hidden and visible sectors, let us also consider the case where we obtain a non-chiral operator. One natural way to generate such an operator in this setup is to consider the contribution of an instanton anti-instanton pair wrapping an orientifold invariant cycle, since we then obtain the four neutral fermionic zero modes that make up the full $\mathcal{N}=1$ superspace measure, and we know that the opposite GSO projections in the Ep-Dp$'$ and $\overline{{\rm E}}$p-Dp$'$ sectors will generate factors with opposite chirality.

We then expect the following terms involving the $\overline{{\rm E}}$p-Dp$'$ sector zero modes in the $\overline{{\rm E}}$p action\footnote{One can rigorously derive such couplings for ${\rm \overline{E}3}$ branes in the toroidal orientifold examples we will consider below.}
\be\label{barEpDp}
L=\overline{\alpha}_V \overline{X}^{(V)} \overline{\beta}_V + \overline{\alpha}_H \overline{X}^{(H)} \overline{\beta}_H ~.
\ee
Heuristically, we also expect the following terms involving interactions between the $\overline{{\rm E}}$p-Ep states
\be\label{barEpEp}
L=(x_-^2-\frac{1}{2})|\varphi|^2+ix_{-\mu}\bar{\chi}\sigma^{\mu}\chi+V
\ee
where $x_-^{\mu}=x_1^{\mu}-x_2^{\mu}$ is the distance in the non-compact four dimensional space between the instanton and anti-instanton, $\chi$ are fermionic zero modes with one end on each the instanton/anti-instanton, and $\varphi$ is a bosonic mode stretching between the instanton and the anti-instanton. $V$ is a potential depending on $\varphi, \chi, \theta$ whose precise form is not important. 
Note that the terms appearing in (\ref{barEpEp}) are analogous to the terms appearing in the corresponding $\overline{{\rm D}}{\rm (p}+4)$-D(p$+4)$ action with $\varphi$ playing the role of a \lq tachyon' in the following sense: about $\varphi=0$, we can think of the above action as describing an instanton anti-instanton pair, while about the minima with $\varphi\ne 0$ (the precise location of these minima depends on the unspecified potential, $V$), the above description breaks down since the branes have annihilated. In particular, the zero mode content and interactions described in (\ref{EpDp}), (\ref{barEpDp}), and (\ref{barEpEp}) are really only an effective description of the physics about $\varphi=0$ since these modes cease to exist about the minima with $\varphi\ne 0$.

Note, however, that for large $x_-^2$ we expect the configuration about $\varphi=0$ to be stable in the sense that it is a local minimum of the $\varphi$ potential. Indeed, for large $x_-^2$, we can integrate out the fields $\varphi$ and $\chi$ and so we expect to have a well-defined semi-classical contribution by the instanton anti-instanton pair to the non-chiral operator generated by integrating over the various Ep-Dp$'$ and ${\rm\overline{E}p}$-Dp$'$ zero modes with the action given by the sum of (\ref{barEpDp}) and (\ref{EpDp}). Physically what is happening is that the only contributions to the operator of interest come from configurations where the instanton and anti-instanton are far apart and hence non-interacting. In particular, we expect to be able to generate a contribution to the K\"ahler potential of the form
\be\label{cont1}
K=M_s^{2(1-d_H-d_V)}e^{-2V_{\Sigma}/g_s}\mathcal{O}_H\overline{\mathcal{O}}_H\mathcal{O}_V\overline{\mathcal{O}_V}
\ee
Note that in writing this formula we have assumed that the instantons and anti-instantons wrap the same cycle. Presumably we could also consider instantons and anti-instantons that wrap different cycles. These pairs of branes would generate contributions to the K\"ahler potential of the form
\be\label{Kcont2}
K=M_s^{2(1-d_H-d_V)}e^{-(V_{\Sigma}+V_{\tilde{\Sigma}})/g_s}\mathcal{O}_H\overline{\tilde{\mathcal{O}}}_H\mathcal{O}_V\overline{\tilde{\mathcal{O}}_V}
\ee
where in general $\mathcal{O}_{H,V}\ne\tilde{\mathcal{O}}_{H,V}$.

Finally, please note that the non-holomorphic operators in (\ref{Kcont2}) could also be generated by single, non-BPS instantons that wrap non-holomorphic, volume minimizing cycles since these instantons also have the requisite four Goldstino zero modes, $\theta, \bar{\theta}$. In fact, this is the generic case.

\subsection{Instanton mediation}

Let us now discuss in some more detail how this setup results in mediation of SUSY breaking from the hidden sector to the visible sector.

The visible sector consists of D-branes on which the MSSM or another supersymmetric extension of the SM is realized. We denote the chiral superfields in this sector as $\Phi_{V,i}$ and its superpotential as $W_V(\Phi_{V,i})$.
The hidden sector is a set of D-branes on which SUSY is broken. Its fields and superpotential are denoted $\Phi_{H_j}$ and $W_H(\Phi_{H,i})$ respectively. Let us focus on the case in which SUSY is broken by (some) non-vanishing F-term vev(s), $F_{\Phi_{H,0}}$---note that this state may simply be metastable with a long lifetime. 

The mediating instantons described above generate an exponentially suppressed perturbation of the superpotential, $W_{H/V}(\Phi_{V,i},\Phi_{H,j})$, coupling the two sectors. The total superpotential reads 

\beq
W=W_V(\Phi_{V,i})+W_H (\Phi_{H,j})+ W_{H/V}(\Phi_{V,i},\Phi_{H,j}) ~.
\eeq 
This superpotential can give rise to a corresponding non-zero F-term vev(s) for some field(s), $F$. Note that in general $F$ {\it need not} be equal $F_{\Phi_{H,0}}$. However, if $W_{H/V}$ is a small perturbation, we expect that $F\sim F_{\Phi_{H,0}}$, and we also expect that the (meta) stability of the SUSY breaking state is not affected. This statement further assumes that some form of mediation---instanton or otherwise---generates masses for the visible sector scalars and that the SUSY breaking sector has all its moduli lifted as well. In this approximation, soft terms arise as follows. We have

\beq
\left\langle {\partial W_H \over \partial \Phi_{H_0}}\right\rangle \neq 0 \ \ \ \ \ \ \ \left\langle {\partial W_{H/V} \over \partial \Phi_{H,V_i}}\right\rangle = \left\langle {\partial W_{V} \over \partial \Phi_{V_i}}\right\rangle= 0 ~.
\eeq
Plugging the non-zero F-term into the superpotential, we see that $W_{H/V}$ can give rise to various SUSY breaking terms. Similarly, plugging the non-zero F-term into the instanton/anti-instanton induced perturbation to the K\"ahler potential described in (\ref{Kcont2}), will yield additional, non-holomorphic SUSY-breaking terms.\footnote{Recall that these terms could also be generated by appropriate non-BPS instantons.}

As a final comment, we note that the instanton induced-operators we have written down are generally non-renormalizable. In particular, the non-renormalizable operators will be suppressed by powers of the string scale, $M_s$. If the F-term vevs are of the scale $F\sim M_s^2$, the suppression by the instanton volume is crucial to obtaining phenomenologically reasonable SUSY breaking scales in the range of several TeV. Of course, it turns out that in many scenarios, including the examples we discuss below, it is possible to have $F\ll M_s^2$. For example, this can happen in D-brane gauge theories that break SUSY dynamically and that have a dynamical scale $\Lambda\ll M_s$ or, for certain values of the moduli, in theories where the SUSY breaking vev is generated by stringy instantons. The mediating instanton then generates an additional suppression with respect to the string scale. Whether the resulting instanton-induced soft terms are an important effect or not depends, as we will see below, on where one sits in the moduli space of the compactification (in particular, in these latter cases, the 4-cycles wrapped by the mediating instantons should be relatively small)

\subsection{Possible soft terms}
In this section we would like to describe more precisely which soft terms we expect to be able to generate via instanton mediation. In order to understand this point, let us first recall the general form of soft terms in renormalizable SUSY gauge theories, of which the MSSM is, of course, an example. For concreteness, consider the following superpotential
\be\label{EgW}
W=\lambda\Phi^3+M\Phi^2+\frac{\tau}{4}W_{\alpha}W^{\alpha} ~,
\ee
where $\Phi$ is shorthand for the various chiral superfields of the theory and $W_{\alpha}$ is a field strength superfield. In general, the holomorphic soft terms are those terms that can be written as higher components of the superfield couplings. In particular, turning on F-term components of $\lambda$ and $M$ results in scalar trilinears $\phi^3$ and scalar bilinears $\phi^2$ called \lq A-' and \lq B-' terms respectively. Turning on an F-term component in $\tau$ generates a gaugino mass.

From the above discussion, it should be clear that instanton mediation can generate both A-terms and B-terms. Indeed, we expect that

\begin{itemize}
\item Instanton mediation generates A-terms when a mediating instanton has zero modes that are charged under the gauge groups of two intersecting $SU(3)$ nodes of the visible sector quiver {\it or} when it has zero modes that transform in the ${\bf \bar{6}}$ (${\bf 6}$) representation of the gauge group of a visible sector $SU(6)$ node that has an antisymmetric (conjugate antisymmetric) tensor representation after orientifolding.

\item Similarly, instanton mediation generates B-terms when one considers the scenarios mentioned in the A-term generation case but with $SU(2)$ nodes instead of $SU(3)$, or $SU(4)$ with an antisymmetric tensor instead of $SU(6)$ with an antisymmetric tensor.
\end{itemize}

Now, taking into account the couplings given in (\ref{Gkincorr2}), we expect that instanton mediation also generates gaugino masses in certain cases.

The remaining soft terms are non-holomorphic. For our purposes, the only interesting non-holomorphic soft terms arise from $\theta^4$ components of the coupling $Z$ in the K\"ahler potential
\be\label{ZK}
K=Z\bar{\Phi}\Phi
\ee
These soft terms are non-holomorphic soft masses, $\bar{\phi}\phi$. It should be obvious from the above discussion that instanton mediation also generates these terms in some cases. In particular, we find that
\begin{itemize}
\item Instanton mediation generates a non-holomorphic mass term for visible sector fields that run between two intersecting $U(1)$ nodes {\it or} for visible sector fields that transform in the antisymmetric, i.e., trivial representation of an $SU(2)$ node after orientifolding.\footnote{As a brief aside, note that these instanton generated non-holomorphic mass terms could be used, for example, to give a mass to the \lq right-handed' scalar neutrino. Furthermore, we should also note that we can obtain exponentially-suppressed non-holomorphic mass terms for the squarks and the other sleptons by taking the corresponding higher-dimensional instanton-generated operators---heuristically of the form $e^{-2V_{\Sigma}/g_s}\frac{F^2}{M_s^{2n}}\bar{\phi}^n\phi^n$--- and contracting $n-1$ pairs of fields or alternatively through diagrams involving A-terms.}
\end{itemize}

Of course, instanton mediation also generically generates higher-dimension terms. However, these terms in the potential are non-renormalizable and hence the corresponding power-law corrections to the 1-PI effective action will not ruin the softness of our mediation mechanism.\footnote{One might also worry about soft terms of the form $\bar{\phi}\phi\phi$ which might generate unacceptable quadratic divergences in the effective action. Such divergences will not arise in our setups due to symmetries of our quivers.}

\subsection{Possible hidden sectors and mediation mechanisms}
Now that we have described the soft terms that we can potentially generate from our mediation mechanism, let us turn our attention to the possible SUSY breaking sectors that we can include in our setup. String compactifications feature a broad array of possible hidden sectors each of which can have either stable or metastable SUSY breaking vacua. For example, we can have:

\begin{itemize}

\item Sectors that realize simple SUSY breaking models without
non-abelian gauge dynamics (such as Polonyi, Fayet or O'Raighfertaigh models). 
\item Hidden sectors that realize an ordinary gauge theory on D-branes with dynamical SUSY breaking. 
\item A hidden sector with anti D-branes.

\end{itemize}

\noindent Later we present explicit examples of the first two types of hidden sectors in the context of toroidal orientifolds where D-instanton mediation generates various soft terms. We leave a discussion involving the third type of hidden sector to future work. Also note that more than one class of hidden sector can be simultaneously present in a given compactification---for the sake of simplicity, however, all our explicit examples below will have a single type of SUSY breaking hidden sector. 

On top of this, more than one mediation mechanisms can act 
at the same time. For example, gravity and gauge mediation\footnote{We are really using these terms imprecisely as catch-all phrases for closed and open string mediation respectively} are always present in the sense that one always has open and closed string exchange between the hidden and visible sectors (although the exchange may be suppressed). Different mechanisms become dominant over certain regions of the moduli space. Similarly, 
various soft terms might get their dominant contribution from different mediation mechanisms. For instance, global symmetries will often prevent perturbative generation of certain soft terms by gauge mediation. These soft terms might instead be generated by D-instantons, and their hierarchy will then encode the relative volumes of the corresponding D-instantons.

\subsection{Relative dominance of mediation mechanisms}
In this section we will give a heuristic sketch of when one can generally expect different mediation mechanisms to become important in a given compactification. From our discussion above, we know that instanton mediation generates soft terms of the form
\be\label{Esoft}
m^2_{\phi}\sim \frac{F^2}{M_s^2}e^{-2V/g_s}, \ \ \ A\sim \frac{F}{M_s}e^{-V'/g_s}, \ \ \ b\sim Fe^{-\tilde{V}/g_s}
\ee
where we have denoted the different instanton volume factors $V$, $V'$, and $\tilde{V}$ to underline the fact that they need not be equal in general.\footnote{We have assumed that $m^2_{\phi}$ is generated by an instanton-anti-instanton pair in \ref{Esoft}.} In fact, if we want to avoid generating $b$ that is too large (a typical problem in minimal forms of gauge mediation), then we would need $\tilde{V}\sim2V, 2V'$.

Now we note that open and closed string mediation is also generically present in string compactifications. Therefore, a natural question is how strong instanton mediation is relative to these other mechanisms and whether it is dominant in some regime. In order to answer this, we first need to have a very basic understanding of the soft scales that are generated by open and closed string mediation. 

Consider closed string mediation first. On general grounds, we expect that mediation from integrating out massive closed string modes generates soft masses at the scale
\be\label{Closedsoft}
m_{\rm cl}^{\rm soft}\sim\frac{F}{M_P}
\ee
where the Planck mass is given by $M_P=\frac{\sqrt{V_Y}}{g_s}M_s$, with $V_Y$ the compactification volume in string units. In fact, (\ref{Closedsoft}) represents an upper bound for massive closed string mediation. Indeed, as argued in \cite{Kachru:2007xp}, this type of mediation can be sequestered by considering compactifications with a hierarchy of length scales. In such cases we obtain
\be\label{Closedsoftseq}
m_{\rm cl}^{\rm soft}\sim e^{-d/R}\frac{F}{M_P}
\ee
where $d\gg R^{-1}$ is the distance between visible and hidden sectors and $R^{-1}$ is, roughly speaking, a typical mass scale of a mediating bulk KK mode. In fact, string theory naturally accommodates such hierarchies of scale due to the warping associated with large numbers of D-branes---we will not, however, consider such effects in the examples we discuss below although such a study would certainly be worthwhile. It should be clear, however, that by considering, e.g., instantons wrapping 4-cycles that have smaller dimensions transverse to a principle dimension of length $\sim d$, we can arrange for instanton mediation to dominate gravity mediation.\footnote{We should also note that anomaly mediation, due to the superconformal anomaly, is also generically present. However, its contributions are suppressed relative to (\ref{Closedsoft}) by various potentially small couplings.}

Next, consider open string mediation. Roughly speaking, such mediation is due to open strings that stretch between the hidden and visible sectors and is characterized by a soft scale
\be\label{Opensoft}
m_{\rm op}^{\rm soft}\sim\frac{g^2}{16\pi^2}\frac{F}{M_g}
\ee
where $M_g$ is a supersymmetric mass associated with the tension of the string
\be\label{Mg}
M_g=d \, M_s
\ee
where $d$ is, again, the distance between the visible and hidden sectors in string units and $g^2/16\pi^2$ is a 1-loop factor of the appropriate D-brane gauge group. These soft terms arise, at least in the simple effective field theory picture (and in a simple form of gauge mediation), from integrating out \lq messenger' fields (i.e., the open string modes we have described) that have acquired non-supersymmetric masses of the form $M_g\pm F$ from their couplings to SUSY breaking fields in the hidden sector. For small couplings, $g$,
we can arrange for gauge mediation to be subdominant to instanton mediation. Note that the naive effective field theory picture of open string mediation presumably breaks down for $d\ge1$ since then string oscillator states become important. 

We should also again note that by our above discussion, various mechanisms may be present at once. As we have emphasized above and will see in greater detail below in our explicit examples, different mediation mechanisms---although present---may not even generate certain terms or only generate suppressed terms of a certain type.

\section{Toroidal orientifolds}

\label{section_toroidal_orientifolds}

For concreteness, in this paper we focus on Type IIB compactifications using toroidal orientifolds. For a comprehensive and clear 
explanation of D-branes and instantons at singularities and their embeddings in toroidal orientifolds we refer
the reader to \cite{Aldazabal:2000sa,Ibanez:2007tu}, to whose notation we adhere. For fast reference, we collect some basic formulas needed 
for our constructions and give some tips that are useful for identifying D-brane instantons producing desired couplings.

We consider six-dimensional factorized tori of the form $T^6 = T^2 \times T^2 \times T^2$. Before orbifolding, the theory on a stack of $n$ D3-branes is ${\mathcal N}=4$ $U(n)$ SYM, wich contains $U(n)$ gauge bosons, four adjoint fermions, and six adjoint real scalars. They transform in the ${\bf 4}$ and ${\bf 6}$ of the $SU(4)$ R-symmetry group, respectively. We quotient by the $\IZ_N$ orbifold\footnote{$\IZ_M \times \IZ_N$ orbifolds can be studied analogously.}, which acts on the fermions through the matrix

\beq
{\rm{\bf R_4}}=\diag(\alpha_N^{a_1},\alpha_N^{a_2},\alpha_N^{a_3},\alpha_N^{a_4}) ~,
\label{ZN_4}
\eeq
with $\alpha_N=e^{i 2 \pi/N}$ and $a_1+a_2+a_3+a_4=0 \mod(N)$. From the action on the ${\bf 4}$ we can easily derive the action on the ${\bf 6}$, which is given by

\beq
{\rm{\bf R_6}}=\diag(\alpha_N^{b_1},\alpha_N^{-b_1},\alpha_N^{b_2},\alpha_N^{-b_2},\alpha_N^{b_3},\alpha_N^{-b_3}) ~,
\label{ZN_6}
\eeq
where $b_1 = a_2 + a_3$, $b_2 = a_1 + a_3$ and $b_3 = a_1 + a_2$. We can combine the scalars into complex coordinates $z_s$ on each $T^2$, with $s=1,2,3$. In terms of these degrees of freedom, the identification that follows from \eref{ZN_6} is $z_s \sim z_s \alpha_N^{b_s}$. In order to preserve SUSY, we must have $b_1+b_2+b_3=0 \mod (N)$. The $\IZ_N$ must act crystalographically on the lattice defining the torus. All possibilities have been classified in \cite{Dixon:1986jc}. Each $T^2$ is defined by

\beq\label{Torusaction}
z_s \sim z_s + r_s \sim z_s + r_s \, \alpha_N ~,
\eeq
where $r_s$ is the corresponding radius. 

In order to completely determine the $\IZ_N$ action, we must also specify how its generator, $\theta$, acts on the Chan-Paton (CP) factors of the various
(Euclidean) D-branes. This action is encoded in a matrix that, for each kind of brane, takes the form

\beq
\gamma_\theta=\diag(\id_{n_0},\alpha \id_{n_1},\ldots,\alpha^{N-1} \id_{n_{N-1}})
\label{gamma_D3} ~,
\eeq
where $\id_{n_i}$ denotes the $n_i$-dimensional identity matrix. 

Since we are interested in orientifolds, due to the possibility of then lifting unwanted extra fermionic zero modes of the instantons in our setups, we further quotient by $\Omega (-1)^{F_L} R_1  R_2 R_3$, with $\Omega$ the orientation reversal on the worldsheet, $F_L$ the left-moving fermion number and $R_s$ the reflection on each plane, $z_s \to - z_s$. As a result, we obtain $64$ O3-planes whose
positions on each $T^2$ are given by

\beq
0, \ \ \ \ \ \ {1\over 2} r_s, \ \ \ \ \ \ {1\over 2} r_s \alpha_N^{1/2}, \ \ \ \ \ \  {1\over 2} r_s \alpha_N
\eeq
Depending on the orbifold action, some of these O-planes may also sit on top of
orbifold fixed points. Furthermore, each O-plane has two possible RR charges. In this paper we will only consider the case in which all O3-planes have negative RR charge (and hence lead to SO and antisymmetric projections of vector and chiral multiplets, respectively).

In any compactification, there are global and local consistency conditions corresponding to cancellation
of untwisted and twisted RR charges, respectively. Due to the presence of the 64 O3-planes,
cancellation of untwisted or global tadpoles reads 

\beq
N_{D3} - N_{\overline{D3}}=32
\label{untwisted_tadpoles}
\eeq
with the net number of other Dp-branes vanishing.\footnote{For simplicity, we limit our discussion to models with only D3 and anti D3-branes in this paper. In fact, our explicit examples do not even contain anti D3-branes.}

When D3-branes sit on top of an orbifold fixed point that does not coincide with an O-plane, twisted or local tadpole cancellation requires

\beq
[\prod_{s=1}^3 2 \sin (\pi k b_s/N)] \Tr \gamma_{\theta^k,3}=0
\label{twisted_tadpoles_orbifold} ~,
\eeq
for each element $\theta^k$, $k=0,\ldots,N-1$, of the orbifold group.

The same expression applies for the case of an anti D3-brane. If the fixed point coincides with an O-plane, the expression changes to

\beq
[\prod_{s=1}^3 2 \sin (\pi k b_s/N)] \Tr \gamma_{\theta^k,3}=4
\label{twisted_tadpoles_orientifold} ~.
\eeq
These two conditions ensure cancellation of gauge anomalies in the corresponding gauge theories. In the discussion above, we take the convention of counting RR charges in the covering space.

There are various possibilities for locating D-branes: they can sit at orbifold fixed points, O-planes or in the bulk. However, they must be in configurations that are symmetric under both the orbifold and orientifold groups.

\subsection{E3-brane instantons}

\label{section_E3_instantons}

Consider an E3-brane in the class of geometries we have described above. The CP matrix will have the general form

\beq
\gamma_{\theta,E3}=\diag(\id_{v_0},\alpha \id_{v_1},\ldots,\alpha^{N-1} \id_{v_{N-1}}) ~.
\label{gamma_E3}
\eeq
Let us take an instanton wrapping $z_s=const$ and intersecting a stack of D3-branes at an orbifold fixed point. In this case, the fermionic zero modes in the E3-D3 sector were computed in \cite{Ibanez:2007tu}. The result is

\beq
\begin{array}{ccl}
a_s \mbox{ even} & \ \ \ \ \ \ &
\sum_{i=0}^{N-1} [(n_i,\bar{v}_{i-{1\over 2} a_s})+(v_i,\bar{n}_{i-{1\over 2} a_s})] \\
a_s \mbox{ odd} & \ \ \ \ \ \ &
\sum_{i=0}^{N-1} [(n_i,\bar{v}_{i-{1\over 2} (a_s+1)})+(v_i,\bar{n}_{i-{1\over 2} (a_s+1)})] 
\end{array}
\label{general_fermionic_zero_modes}
\eeq

Furthermore, the instanton has four additional neutral fermionic zero modes due to the accidental $\mathcal{N}=2$ SUSY of the E3-E3 sector.  As discussed above, a simple way of projecting out the two accidental neutral zero modes is by placing the E3-brane on top of an orientifold with an $O(1)$ projection. Since the orientifold acts by conjugating the CP matrix, this determines

\beq
\gamma_{\theta,E3}=1 ~,
\label{CP_E3}
\eeq
i.e. $v_0=1$ and $v_i=0$ for $i\neq 0$. Plugging this into \eref{general_fermionic_zero_modes}, we conclude that the fermionic zero modes connecting such an instanton to the D3-branes transform as follows under the gauge symmetries of the quiver

\beq
\begin{array}{cclcl}
a_s \mbox{ even} & \ \ \ \ \ \ & \fund_{{1\over 2} a_s} & \ \ \ \ & \antifund_{-{1\over 2} a_s} \\
a_s \mbox{ odd} & \ \ \ \ \ \ & \fund_{{1\over 2} (a_s+1)} & \ \ \ \ & \antifund_{-{1\over 2} (a_s+1)} 
\end{array}
\label{E3_D3_2}
\eeq

The last remaining ingredient in our description of the E3-branes is to give an explicit embedding of their worldvolumes in our orientifolded $T^6/ \mathbb{Z}_N$ compactification. In order to do this, let us first go to the covering space of the orbifold, the orientifold, and the torus. The total covering space is $\mathbb{C}^3$, and we will consider E3-branes wrapping divisors
\be\label{E3cycletotal}
t_1\frac{z_1}{r_1}+t_2\frac{z_2}{r_2}+t_3\frac{z_3}{r_3}=v
\ee
where we have normalized by the various radii $r_s$ of the $T^2_s$ and have included arbitrary complex coefficients $t_s$ and $v$. However, in order for (\ref{E3cycletotal}) to represent a divisor wrapped by an E3-brane, it must be compatible with the various geometric projections---let us now run through the list and see how they restrict the complex surfaces wrapped by the E3's.

We define compatibility with the $T^6$ projection given in (\ref{Torusaction}) to mean that the E3-brane wraps a non-trivial closed curve, i.e., an element of the 4-homology group, $H_4(T^6)$. This fact requires
\be\label{Toruscondition}
t_s=n_s, \ \ \ n_s\in\mathbb{Z}, \ \forall \, s=\{1,2,3\}
\ee
with the $n_s$ relatively prime. One easy way to see this is to note that a basis of mutually holomorphic embeddings is given by
\be\label{H4basis}
\Sigma_s=\{z\in T^6| \ z_s=0\}
\ee
Hence, the integers $n_s$ give the wrapping numbers with respect to this basis. For all the $n_s$'s coprime, the wrapping numbers are simply
\be\label{Wrappingnums}
\omega(\Sigma_s)=|n_\nu|. 
\ee
The volume of a curve, $\Sigma$, wrapped by some E3-brane is then given by
\be\label{VolcurvT6}
V(\Sigma)=\sqrt{\sum_s \omega(\Sigma_s)^2V(\Sigma_s)^2}
\ee

Next, let us consider the action of the orbifold group on E3-branes. The geometric action was given in the discussion immediately following (\ref{ZN_6}) and is reproduced below
\be
z_s \sim z_s\alpha_N^{b_s}
\ee
where $\mathcal{N}=1$ SUSY requires $b_1+b_2+b_3=0 \mod (N)$. The cycle wrapped by the instanton may or may not be invariant under the orbifold
group. For generic cases in which all the $b_s$'s are different, the only invariant cycles are given by the $\Sigma_s$. In the examples below, we will take the orbifold action to be $b=(-1,-1,2)\sim(2,2,2)$ for $N=3$ and hence any cycle of the form
\be\label{Orbinvarcycles}
n_1\frac{z_1}{r_1}+n_2\frac{z_2}{r_2}+n_3\frac{z_3}{r_3}=v
\ee
is orbifold invariant so long as
\be\label{Orbinvarv1}
v=\alpha_3v
\ee
up to identifications (i.e., up to the toroidal shifts in the $z_i$). Note that for $b=(-1,-1,2)$ and $N\ne 3$, however, the only orbifold invariant cycles are given by $n_1,n_2\in\mathbb{Z}, n_3=0$ or $ n_1,n_2=0$, $n_3=1$.

In general, for non-invariant cycles, we must include $N-1$ additional image E3-branes. We can further
divide the orbifold non-invariant cycles into two groups. On the one hand, we have those cycles that go through
orbifold fixed points. In this case, the image E3-branes intersect, giving rise to additional 
neutral fermionic zero modes that must be lifted in order for the E3-brane to contribute to the action. This can be achieved, for example, in compactifications with fluxes. On the other hand, we have non-invariant cycles that do not pass through orbifold fixed points. In this case, the E3-brane images  
are spatially separated, we do not get additional zero modes and a superpotential is generated.

Finally, let us consider the orientifold action. Since we want an $O(1)$ instanton (at least for generating holomorphic soft terms), the cycle wrapped by the E3-brane must be mapped to itself under the geometric part of the orientifold action
\be\label{Ofoldaction}
z_s\to -z_s
\ee
Plugging this action into (\ref{E3cycletotal}), we see that the cycle is invariant if and only if
\be\label{Ofoldinvariant}
v=-v
\ee
up to identifications.

As an aside, note that the volumes of the orientifold and orbifold invariant cycles are then given in the quotient space by (\ref{VolcurvT6}) divided by a numerical factor
\be
V(\Sigma)|_{\rm quot}=\frac{1}{(4N)^3}\sqrt{\sum_i\omega(\Sigma_i)^2V(\Sigma_i)^2}
\label{volume_invariant}
\ee
where the factor of $4^3$ is due to the orientifold and the factor $N^3$ is due to the orbifold. The volume of orientifold invariant cycles that are not invariant under the orbifold group are given by \eref{volume_invariant}, but without the $1/N^3$ factor.

Now, given an instanton wrapping a particular cycle characterized by some coefficients, $n_s$, subject to the constraints just discussed, we would like to understand which operators are generated when the instanton intersects a spacefilling D3-brane. By simple worldsheet CFT arguments, the resulting R-sector fermionic zero modes must have Dirichlet-Dirichlet boundary conditions and hence are in the complex dimension transverse to both the E3 and the D3. For concreteness, then, we see that an E3-brane wrapping $z_s=const$ must couple to the adjoint, $\Phi^s$, of the D3-brane. If this D3-brane sits at an orbifold fixed point, then the E3-D3 zero modes couple to the bifundamental $X^{s}_{ij}$ determined by \eref{E3_D3_2}.

More generally, instantons with worldvolume given by a linear combination with various coefficients, $n_s$, non-zero generate couplings to operators made out of the corresponding combinations of fields. In other words, the orientation of the instanton selects the type of bifundamentals that form the operator. After performing the zero mode integral, the orientation of the instanton then picks out a chiral gauge invariant of the same orientation.

As a final point, let us consider more specifically the possible forms of the cycles wrapped by {\it mediating} instantons. Given a hidden and a visible sector, more than one instanton can connect them. For simplicity, let us consider the situation in which one of the sectors is located at the origin, as will happen in our examples below. This implies that we can set $v=0$ in \eref{E3cycletotal}. The position of the other sector is $(h_1 \, r_1,h_2 \, r_2,h_3 \, r_3)$, with $h_s \in \IC$. As explained above, we can discard orbifold non-invariant 4-cycles that go through fixed points. In the absence of a mechanism that lifts the additional zero modes (as is the case in our examples), they do not generate mediating interactions. 

For concreteness, let us now assume the orbifold form $b=(-1,-1,2)\sim(2,2,2)$ for $N= 3$. For setups with $(h_1,h_2,h_3)\sim(0,0,h_3^0)$---where \lq$\sim$' means, \lq up to identifications of the $T^6$ action'---we see that the orbifold invariant cycles going through both the visible and hidden sectors are given by
\be
n_1h_1+n_2h_2+n_3h_3=0
\ee
where we generate solutions $n_i$ by substituting $h_{1,2}=0, h_3=h_3^0$ and also substituting values related to these by the action of the $T^6$. If all the $h_i\ne 0$, then a similar discussion applies.

Any solution to this equation with $n_1, n_2, n_3 \in \IZ$ defines a cycle wrapped by a mediating instanton. In general, there is more than one such solution. In practice, due to the exponential suppression, we are only interested in the solutions with the smallest volumes.

\subsection{The $\IZ_3$ orientifold}

\label{section_Z3_orientifold}

A simple way to achieve three generations and generate a crude zeroth order approximation to the Standard Model in the context of D-brane at singularities is via a $\IZ_3$ orientifold. Consequently, we will base our explicit examples on this case, keeping in mind that other geometries allow for more realistic visible sectors and interesting hidden sectors (see for example \cite{Aldazabal:2000sa,Berenstein:2001nk,Verlinde:2005jr,Wijnholt:2007vn}). For later reference, we devote this subsection to a more detailed presentation of the $\IZ_3$ case. A similar general discussion of some of the results for the $\IZ_3$ orbifold appears in \cite{Kakushadze:2002fa}.

We take the (SUSY) orbifold action on the fermions to be given by $(a_1,a_2,a_3,a_4)=(1,1,-2,0)$. From this data, we determine the action on the three complex planes to be given by $(b_1,b_2,b_3)=(-1,-1,2)$
. This model contains $27$ orbifold fixed points. Their positions are $(z_1,z_2,z_3)$, with $z_s=0,{1\over \sqrt{3}}e^{i\pi/6},{1\over \sqrt{3}}e^{-i\pi/6}$. In what follows, we refer to these points as $0$, $+1$
and $-1$ respectively. Out of the 64 O3-planes, only the one at the origin, $(0,0,0)$, coincides with an orbifold fixed point. Note that the orientifold action leaves $z_s=0$ invariant and interchanges $z_s=\pm 1$. 

Given this discussion, we can write down the general solutions to the twisted tadpole cancellation equations \eref{twisted_tadpoles_orbifold} and \eref{twisted_tadpoles_orientifold} at orbifold fixed points. Since the orbifold fixed points at $(z_1,z_2,z_3)\neq (0,0,0)$ do not coincide with an O-plane, they must satisfy \eref{twisted_tadpoles_orbifold}. The general solution reads

\beq
\gamma_\theta=\diag(\id_N,\alpha \id_N,\alpha^2 \id_N)
\label{gamma_orbifold_Z3}
\eeq
The resulting gauge theory has a $U(N)\times U(N) \times U(N)$ gauge group and matter content

\beq
\begin{array}{c|cc}
\ \ \ \ \ \ \ \ \ & \ \ \ & U(N) \times U(N) \times U(N) \\ \hline
X^s_{01} & & (\fund,\bar{\fund},1) \\
X^s_{12} & & (1,\fund,\bar{\fund}) \\
X^s_{20} & & (\bar{\fund},1,\fund)
\end{array} 
\label{spectrum_orbifold_Z3}
\eeq
with $s=1,2,3$ and with the subindices indicating the gauge groups under which bifundamental fields transform. The overall $U(1)$ is anomaly free but decouples since all fields are neutral under it. The other two linear combinations of $U(1)$'s have mixed anomalies and become massive via the $B\wedge F$ couplings of the Green-Schwarz mechanism. The superpotential is

\beq
W=\epsilon_{stu} X^s_{01} X^t_{12 }X^u_{20} ~,
\label{W_orbifold_Z3}
\eeq
where we have suppressed color indices for simplicity.

The origin $(z_1,z_2,z_3)=(0,0,0)$ is an orientifold singularity and hence \eref{twisted_tadpoles_orientifold} holds. The most general solution is

\beq
\gamma_\theta=\diag(\id_N,\alpha \id_{N+4},\alpha^2 \id_{N+4})
\label{gamma_orientifold_Z3}
\eeq
This results in a gauge theory with an $SO(N)\times U(N+4)=SO(N)\times SU(N+4)\times U(1)$ gauge group, with matter transforming as 

\beq
\begin{array}{c|cc}
\ \ \ \ \ \ \ \ \ & \ \ \ & SO(N) \times SU(N+4) \times U(1) \\ \hline
\bar{Q}^s & & (\fund,\antifund)_{-1} \\
A^s & & (1,\symm)_2 
\end{array} 
\label{spectrum_orientifold_Z3}
\eeq
where $s=1,2,3$. There are mixed anomalies and the $U(1)$ factor becomes massive due to $B\wedge F$ couplings. The superpotential is given by

\beq
W=\epsilon_{stu} \bar{Q}^s A^t \bar{Q}^u~.
\label{W_orientifold_Z3}
\eeq

The gauge theory of $N$ D3-branes sitting on an O-plane that is not at an orbifold fixed point is ${\mathcal N}=4$ SYM with $SO(N)$ gauge group and three antisymmetric chiral fields $A^n$. The superpotential in this case is

\beq
W=\epsilon_{stu} A^s A^t A^u~.
\label{W_N4}
\eeq

Let us now consider the couplings generated by an E3-brane instanton. Equation \eref{E3_D3_2} gives the fermionic zero modes between the instanton and the (fractional) D3-branes in either the hidden or visible sectors.
Specializing to the case at hand, we get fermionic zero modes transforming in the $\fund_1$ and $\antifund_2$ representations for instantons wrapping $z_s=0$, for $s=1,2,3$. The subindices $1$ and $2$ of the representations denote the quiver nodes associated with the $\alpha$ and $\alpha^2$ blocks of the CP matrix. As a result, any instanton corresponding to a general linear combination of the form \eref{E3cycletotal}, generates a coupling of the form

\beq
W_{inst}=\det X_{12} \ \ \ \ \ \ \mbox{or} \ \ \ \ \ \ W_{inst}=\sqrt{\det A} ~,
\eeq
where the second possibility corresponds to the case in which nodes 1 and 2 are identified by the orientifold. $X_{12}$ and $A$ in the previous expressions correspond to the linear combinations of $X_{12}^s$ and $A^s$ that are determined by \eref{E3cycletotal}. We see that the instanton generates couplings involving only 
fields connecting nodes 1 and 2 (before orientifolding) in both hidden and visible sectors. For this reason, we are interested in hidden sectors in which operators made out of $X_{12}$ or $A$ have a non-zero F-term. 

Similar reasoning applies to the case of D3-branes on O-planes that are not orbifold fixed points. In fact, we can simply take the general expressions for $\IZ_N$ orbifolds and set $N=0$. The instanton generated superpotential is, once again,

\beq
W_{inst}=\sqrt{\det A} ~,
\eeq
with $A$ the linear combination of antisymmetrics associated with the specific instanton embedding.

For D3-branes over orientifolds (either on orbifold singularities or not), the instanton generated coupling is non-zero only when the $SU(n)$ gauge group has even $n$, since the determinant of an odd order antisymmetric matrix vanishes. This is true if the corresponding sector is either the visible or the hidden sector. For even $n$, $\det A$ is a perfect square, and the following way of writing its square root is sometimes convenient
\beq
\sqrt{\det A}=\frac{1}{2^n \, n!}\epsilon^{a_1 \ldots a_n} A_{a_1 a_2} \ldots A_{a_{n-1}a_n}:={\rm Pf}(A)~.
\eeq

\section{Examples}

\label{section_examples}

In this section we present explicit models of D-instanton mediation involving different classes of hidden sectors. 
Our goal is to provide illustrative examples that show how D-brane instantons can communicate SUSY breaking rather than constructing models with fully realistic visible sectors. In sections 5 and \ref{section_more_ingredients} we discuss some possible directions for constructing more elaborate models.

\subsection{Dynamical SUSY breaking hidden sector}
In this subsection we present a model in which the hidden sector breaks SUSY via some non-abelian gauge dynamics. 

We consider the $\IZ_3$ orientifold of section \ref{section_Z3_orientifold}. We place the hidden sector on top of the orientifold singularity at the origin, with CP matrix given by taking $N=0$ in \eref{gamma_orientifold_Z3}, i.e.

\beq
\gamma_{\theta,3}=\diag(\alpha \id_4,\alpha^2 \id_4)
\eeq 
The resulting gauge theory has an $SU(4)\times U(1)$ gauge symmetry and 
three fields $A^i$ in the $\symm_2$. The $\bar{Q}^s$'s
from \eref{spectrum_orientifold_Z3} are absent and hence there is no tree-level superpotential.

Let us forget about the $U(1)$ factor for the moment. This model has been considered in \cite{Kakushadze:2002fa}. It can be alternatively viewed as an $SO(6)$ gauge theory with three flavors of quarks in the vector representation. The theory is strongly coupled in the IR.\footnote{The beta function for the inverse coupling, computed in both pictures, is equal to 9.} The low energy theory has two physically inequivalent phase branches \cite{Intriligator:1995id}. On one of them, a non-perturbative superpotential is generated

\beq
W_{H,np}=8 {\Lambda^9 \over \det M} ~,
\label{Wnp_SU4}
\eeq
with $M$, the symmetric (over the $SU(3)$ flavor indices) meson matrix. SUSY is broken on this branch with runaway along $M$. The previous expression can be written in terms of the antisymmetrics of $SU(4)$. 
For diagonal $M$, we obtain 

\be
W_{H,np}=8 {\Lambda^9 \over \prod_s {\rm Pf}(A^s)} ~.
\label{Wnp_SU_A}
\ee
Reintroducing the $U(1)$ factor, the scalar potential contains a D-term contribution of the form

\beq
V_{U(1)}={1\over \lambda}(\sum_s 2 |A^s|^2-\xi)^2 ~,
\label{V_D}
\eeq
where $\xi$ is a {\it dynamical} FI term, which is related by SUSY to the $B \wedge F$ coupling that makes the $U(1)$ massive. If there is a mechanism stabilizing all K\"ahler moduli, $\xi$ becomes a fixed parameter and \eref{V_D} cures the runaway, producing a non-SUSY vacuum. For the purpose of illustration, we content ourselves with the fact that, regardless of whether the runaway is stabilized or not, there is a non-vanishing $F_M$ on this branch.\footnote{This model has also been considered in \cite{Ibanez:2007tu} in connection with SUSY breaking, although with a completely different approach. In that case, a mass term for the antisymmetrics is generated by a D-brane instanton and SUSY breaking results from its interplay with a fixed non-vanishing $\xi$.} The second branch has no non-perturbative superpotential and a quantum moduli space of vacua. The theory confines without chiral symmetry breaking. Since there is no non-perturbative superpotential, SUSY is not broken on this branch.

Our visible sector is a trinification model, a simple extension of the SM that has been investigated in the model building literature \cite{trinification}. We place it at $(1,0,0)$, with its orientifold image at $(-1,0,0)$. Its CP matrix corresponds to taking $N=3$ in \eref{gamma_orbifold_Z3}. We get

\beq
\gamma_{\theta,3}=\diag(\id_3,\alpha \id_3,\alpha^2 \id_3)
\eeq 
This produces an $SU(3) \times SU(3) \times SU(3)$ gauge theory with

\beq
\begin{array}{c|cc}
\ \ \ \ \ \ \ \ \ & \ \ \ & SU(3) \times SU(3) \times SU(3) \\ \hline
X^s_{01} & & (\fund,\bar{\fund},1) \\
X^s_{12} & & (1,\fund,\bar{\fund}) \\
X^s_{20} & & (\bar{\fund},1,\fund)
\end{array} 
\label{spectrum_trinification}
\eeq
with $i=1,2,3$ and superpotential is

\beq
W_V=\epsilon_{stu} X^s_{01} X^t_{12} X^u_{20} ~,
\label{W_trinification}
\eeq
The model is not fully realistic. For example, it does not contain the higgs fields that are necessary to break two of the $SU(3)$'s down to $SU(2) \times U(1)$.

Assuming $r_1\sim r_2 \sim r_3$, the two smallest-volume mediating instantons wrap $z_2=0$ and $z_3=0$. There are additional orbifold invariant 4-cycles connecting the hidden and visible sectors, but contributions from instantons wrapping these cycles are highly suppressed since they have larger volume. From \eref{W_orbifold_Z3} and \eref{W_orientifold_Z3}, our leading-order mediating instantons generate the following superpotential

\beq
W_{H/V}=e^{-V_{\Sigma_2}/g_s} \sqrt{\det A^2} \det X_{12}^2+ e^{-V_{\Sigma_3}/g_s} \sqrt{\det A^3} \det X_{12}^3~.
\label{WHV_SU4}
\eeq
where the orientation of the mediating instantons projects onto gauge invariants of the SUSY breaking hidden sector theory that acquire F-term vevs on the branch with the non-perturbative superpotential given in (\ref{Wnp_SU_A}). Therefore, \eref{WHV_SU4} gives rise to the following A-terms 

\beq
V_{soft}=e^{-V_{\Sigma_2}/g_s} F_{{\rm Pf}(A^2)}^* \det X_{12}^2 |_{\theta=0}+ e^{-V_{\Sigma_3}/g_s} F_{{\rm Pf}(A^3)}^* \det X_{12}^3 |_{\theta=0}+ c.c. ~,
\label{V_soft_SU4}
\eeq
where specializing for $\theta=0$ indicates that we take the scalar component of the visible sector chiral superfields. For simplicity, in this section and the next one, we ommit obvious powers of $M_s$ and the dynamical scale $\Lambda$ of the hidden sector, which are necessary for expressions to have the correct dimensionality. Note that these A-terms correspond to couplings between the Higgs fields and the sleptons. Analogous Yukawa couplings are also generated non-perturbatively by other D-brane instantons (see discussion below).

Our hidden sector and visible sector (plus image) involve 26 D3-branes. We can cancel untwisted tadpoles \eref{untwisted_tadpoles} without spoiling the nice features of our model by placing the 6 additional D3-branes in sets of two over the O3-plane at $({1\over 2} r_1,{1\over 2} r_2,{1\over 2} r_2)$ and its two $\IZ_3$ images.

\subsection{Polonyi hidden sector}

Another exciting direction is the possibility of having a simple field theory hidden
sector that breaks SUSY without involving non-abelian gauge dynamics, along the lines 
of \cite{Aharony:2007db}.\footnote{In a similar spirit, another realization of a Polonyi model involving D-brane instantons appears in \cite{Cvetic:2007qj}.} A remarkably simple possibility is to engineer a Polonyi model. This construction is very general and can easily be part of more complicated setups. Because of this, we consider it deserves to be discussed first, independently of the details of the full compact model.

\subsubsection{Engineering a Polonyi model}

\begin{figure}
\begin{center}
\psfrag{y1}[cc][][.7]{$z_1=0$}
\includegraphics[width=5cm]{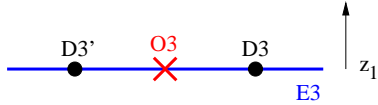}
\caption{The basic configuration realizing a Polonyi model. It consists of an O3-plane and a D3-brane away from it, connected by a finite size E3-brane with $O(1)$ CP projection.}
\label{Polonyi_1}
\end{center}
\end{figure}

The configuration we want to consider consists of an O3-plane and a single D3-brane separated from it, with an E3-brane connecting them. Without loss of generality, we can assume that the E3-brane wraps the $z_1=0$ cycle.\footnote{It is straightforward to
extend our argument to the case in which the instanton is defined by some linear combination of the of the form \eref{E3cycletotal}.} The setup is sketched in \fref{Polonyi_1}, where we have also included the image D3-brane.

The gauge theory on the D3-brane is ${\mathcal N}=4$ $U(1)$ SYM with three chiral superfields $\Phi^s$. The $\Phi^s$ transform in the ``adjoint" representation, which is trivial for $U(1)$ (i.e. they are neutral fields). As a result, the beta function for the gauge coupling is zero and we can tune the gauge coupling to be arbitrarily small. In addition, the ${\cal N}=4$ superpotential \eref{W_N4} vanishes.

If the CP projection on the E3-brane is $O(1)$, it induces a coupling

\beq
W=e^{-V_{\Sigma_1}/g_s}\, \Phi^1 ~.
\label{W_Polonyi_1}
\eeq
This is precisely a Polonyi model superpotential and SUSY is broken by $F_{\Phi_1}\neq 0$. Of course, other instantons contribute to the superpotential. However, by considering the effect of only the $z_1$ instanton, we are implicitly assuming that $r_1 \gg r_2, r_3$. As usual in Polonyi models, $\Phi_1$ is a classically flat direction. Its stability depends on the details of the full model. The same comments apply to the construction we present below. This is a question that certainly deserves more study in our concrete setups.

\begin{figure}
\begin{center}
\psfrag{y1}[cc][][.7]{$z_1=0$}
\includegraphics[width=5cm]{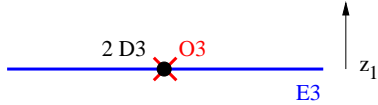}
\caption{A Polonyi model is also obtained when two D3-branes sit on top of an O3-plane. A finite size E3-brane with $O(1)$ CP projection generates the superpotential.}
\label{Polonyi_2}
\end{center}
\end{figure}

Let us now investigate what happens if we collapse the D3 and D3$'$ on top of the O3-plane. To avoid a chiral theory on the D3-branes, we further assume that the O3-plane is not at an orbifold fixed point.
The configuration is shown in \fref{Polonyi_2}.  

The resulting gauge theory is ${\cal N}=4$ $SO(2)$
SYM, with three chiral superfields $A^s$ in the antisymmetric representation. Once again, the beta function for the gauge coupling vanishes and can thus be tuned to any desired value.\footnote{This can be understood as follows. The antisymmetric representation of $SO(N)$ is the same as the adjoint representation. $C(adj)=(N-2)$ for $SO(N)$, and hence vanishes in this case. As a result, the beta function also vanishes. An equivalent way of thinking about the beta function is that $SO(2)=U(1)$ and the antisymmetric representation corresponds to a neutral field.} As before, we exploit this fact to make the gauge coupling small so it can be neglected. The antisymmetric representation of $SO(2)$ is trivial and the $A^s$ have the general form

\beq
A^s_{ab} = \phi_s \, \epsilon_{ab} ~.
\eeq
Then, the ${\cal N}=4$ superpotential \eref{W_N4} vanishes. If the E3-brane has an $O(1)$ CP projection, it generates a coupling
\beq
W=e^{-V_{\Sigma_1}/g_s}\, \sqrt{\det A^1}= e^{-V_{\Sigma_1}/g_s} \, \phi_1
\label{W_Polonyi_2}
\eeq
A square root appears in \eref{W_Polonyi_2} as opposed to
\eref{W_Polonyi_1} because, in this case, the E3-D3 and D3-E3 fermionic zero modes are identified by the orientifold projection.
\footnote{It is interesting to consider what happens for a single D3-brane on top of the O-plane. The coupling \eref{cubic_coupling_instanton_orientifold} vanishes identically due to anti-commutativity of $\alpha$ which, in this case, is one dimensional.} 

In our opinion, this simple realization of a Polonyi hidden sector is
interesting in its own right, independently of which mechanism mediates SUSY breaking. 

Notice that the superpotential terms generated in this section are not perturbatively forbidden by any $U(1)$ symmetry. The appearance of a non-perturbative superpotential determined by the zero of the E-brane embedding is an example of Ganor's zeros \cite{Bershadsky:1996gx,Ganor:1996pe}. In \cite{Bershadsky:1996gx} the ADS superpotential of $N_f=N_c-1$ SQCD was generated along these lines. In that case, the effect is due to a gauge theory instanton since the flavor D7-branes and the E3-brane are wrapped over the same 4-cycle. In our setup, D7-branes are not present and the effect is purely stringy. 
In section 6, we discuss similar operators in more general models.

\subsubsection{A full model: visible sector and mediation}

We now use a hidden sector of the type just described as a part of a simple compactification. We consider the $\IZ_3$ toroidal orientifold of section \ref{section_Z3_orientifold}. In this case, we will use D-brane instantons not only for mediating, but also for generating SUSY breaking. This fact gives rise to constraints on the relative sizes of the tori that are necessary for the model to work.

We engineer a Polonyi hidden sector by placing two D3-branes on top of the O3-plane at $({1\over 2} r_1,{1\over 2} r_2,{1\over 2} r_3)$ (plus its two $\IZ_3$ images). A Polonyi superpotential 

\beq
W=e^{-V_{\Sigma_1}/g_s}\phi_1 
\eeq
is generated by the E3-brane at $z_1/r_1=1/2$. This term is the dominant contribution to the superpotential involving hidden sector fields provided that $r_1 \gg r_2,r_3$ since this condition guarantees that $V(\Sigma_1)\ll V(\Sigma_{2,3})$. Notice that although the cycle wrapped by this instanton is not invariant under the orbifold group, it does not intersect its images and hence there are no extra zero modes. Following the discussion in section \ref{section_E3_instantons}, it generates a non-vanishing contribution.

Next, let us think about where to locate the visible sector. Note that in order not to spoil the generation of the Polonyi superpotential, the instanton wrapping $z_1/r_2=1/2$ must not intersect the visible sector. A particularly simple choice, then, is to locate the visible sector on top of the orientifold singularity at the origin, $(0,0,0)$. As our visible sector, we will choose an interesting GUT-like model that can be engineered as follows \cite{Ibanez:2007tu}. We take $N=2$ in \eref{gamma_orientifold_Z3}

\beq
\gamma_{\theta,3}=\diag(\id_2,\alpha \id_6,\alpha^2 \id_6) ~.
\label{CP_GUT_like}
\eeq
This gives rise to a theory with $U(6) \times O(2)$ gauge group and chiral multiplets transforming as

\beq
3 \, (\overline{15},0)+3 \, (6,+1)+3 \, (6,-1) ~.
\eeq
Under the $SU(5)$ subgroup of $U(6)$, these representations decompose as $\overline{15}=\overline{10}+\bar{5}$ and $6=5+1$, giving rise to three SM generations
$(\overline{10}+5)$ and three sets of higgs fields $(5+\bar{5})$.\footnote{The presence of a couple of copies of each
MSSM or SUSY GUT higgs, although not unavoidable, is a usual feature in D-brane realizations.} As it stands, this model is not fully realistic since it does not contain the Higgs field necessary for breaking the GUT group.

Let us now consider the mediating instantons. In order to couple to $F_{\phi_1}$, the embedding equation of a mediating instanton must involve $z_1$---i.e., the orientation of the instanton must project onto the SUSY breaking part of the hidden sector. However, since the hidden sector is located at the origin, the equation must also involve $z_2$ and/or $z_3$. Notice that this geometric fact immediately implies that the mediation term will be a small perturbation of the Polonyi superpotential, since an equation involving $z_{2,3}$ requires that the instanton wrapping numbers on the much larger cycles $\Sigma_{2,3}$ cannot both be zero. Hence, a simple choice for a cycle wrapped by a mediating instanton is to take
\be
z_1/r_1-z_2/r_2=0
\ee
with volume
\be
V(\Sigma_{(1,-1,0)})=\frac{1}{12^3}\sqrt{V(\Sigma_2)^2+V(\Sigma_1)^2}=\frac{1}{12^3}\Big(V(\Sigma_2)+V(\Sigma_1)\cdot\frac{V(\Sigma_1)}{2V(\Sigma_2)}+...\Big)
\ee
where we explicitly see that the mediating instanton will have large volume compared to the Polonyi instanton since $V(\Sigma_2)\gg V(\Sigma_1)$. 

What about additional instantons connecting the two sectors? In general, the answer is quite complicated but can be worked out in detail. However, for simplicity, we will further assume that
\be
V(\Sigma_1)\epsilon_{2,3}\gg 1
\label{extra_suppresion}
\ee
where
\be
\epsilon_{2,3}=\frac{V(\Sigma_1)}{V(\Sigma_{2,3})}\ll 1
\ee
Note that \eref{extra_suppresion} is assumed in addition to the assumption that $V(\Sigma_1)\ll V(\Sigma_{2,3})$. We can explain the motivation for \eref{extra_suppresion} in a general context. Consider 
two instantons with comparable volumes $V\sim V'$. The relative suppression of their contributions 
is given by $e^{-V'}/e^{-V}$. We see that small differences in the volume are exponentially enhanced. Equation \eref{extra_suppresion} amounts to requesting that $e^{-V'}/e^{-V} \ll 1$.
Under these conditions, the four leading-order mediating instantons in this approximation wrap
\bea
\frac{1}{r_1}z_1\pm\frac{1}{r_2}z_2=0 \nonumber \\ \frac{1}{r_1}z_1\pm\frac{1}{r_3}z_3=0
\eea
Thus, the superpotential from each of these mediating instantons is
\be
W_{H/V}^{\pm 2,\pm3}\sim e^{-V_{\Sigma_{\pm 2,\pm3}}/g_s} \left({\phi_1 \over r_1}\pm{\phi_{2,3}\over r_{2,3}}\right) \sqrt{\det \left({A_1 \over r_1}\pm{A_{2,3}\over r_{2,3}}\right)}
\label{WHV_Polonyi}
\eeq
where $\bar{A}_i$ are the three $(\overline{15},0)$ fields in the visible sector. While $\phi_1/r_1 \ll \phi_2/r_2$ in the hidden sector piece, it is only $\phi_1$ that gets a non-vanishing F-term and contributes to the soft terms. We get the following A-terms from each of the instantons

\beq
V_{soft} \sim  e^{-(V_{\Sigma_1}+V_{\Sigma_{\pm 2,\pm3}})/g_s} \, {F^*_{\phi_1} \over r_1} \, \epsilon^{abcdef}\tilde{A}_{ab}\tilde{A}_{cd}\tilde{A}_{ef}|_{\theta=0}+c.c. ~,
\label{soft_Polonyi}
\eeq
where the cycle wrapped by the Polonyi instanton is $\Sigma_1$ and where we have defined $\tilde{A}=(A_1/r_1\pm A_{2,3}/r_{2,3})$. For simplicity, we have omitted an obvious $r_1$ and $r_{2,3}$ dependent normalization of \eref{WHV_Polonyi} and \eref{soft_Polonyi}. 
As an aside, note that these A-terms contain couplings between the Higgs fields and the U-type squarks. The corresponding Yukawa couplings are also generated by D-instantons (see discussion below).

The configuration is still missing 12 D3-branes in order to cancel untwisted tadpoles. A simple way of completing the model without spoiling the features we have just discussed is by placing 6 D3-branes
with CP matrix

\beq
\gamma_{\theta,3}=\diag(\id_2,\alpha \id_2,\alpha^2 \id_2)
\eeq
at each of the $(\pm 1,0,0)$ orbifold fixed points.

\section{Phenomenology and instanton orientation}

The visible sectors we have discussed in the models above are  
deliberately simple and, as a result, unrealistic. For example, as we have  
mentioned, we do not even have all the Higgs fields necessary to break  
to the SM gauge group. However, motivated by the fact that our  
theories contain three generations of matter with various Yukawa  
couplings (among them the Yukawa couplings of the MSSM) and noting that our instantons generate various A-terms for  
the visible fields, we are led to ask a very simple question: are the  
A-term matrices, ${\bf A}_i$, and Yukawa coupling matrices, ${\bf Y}_i 
$, aligned? The main phenomenological motivation for this question is  
that alignment of these matrices guarantees suppression of potentially  
troublesome Flavor Changing Neutral Currents (FCNCs) contributions from the A-terms. In particular,  
if ${\bf A}_i\sim k_i{\bf Y}_i$ we say the matrices are aligned. This  
implies that the A-term contributions to the FCNC processes  
responsible for reactions like $K^0\to\bar{K}^0$ are highly suppressed (though not absent).

Before proceeding, we should make two clarifying points. First, even  
if we can align the A-terms and Yukawas, we should emphasize that  
there are still other soft terms that could generate FCNCs, like the  
non-holomorphic part of the squark mass matrix. Since we have only  
discussed non-holomorphic mass generation by instantons in the case of  
squarks and sleptons charged under abelian symmetries (note however footnote 8), we will simply assume that  
the physics responsible for the non-holomorphic squark mass generation  
in the examples is flavor blind. Finally, let us also point out that  
demanding ${\bf A}_i\sim k_i {\bf Y}_i$ is generally a sufficient but not  
necessary condition for suppressing FCNC contributions from A-terms. Indeed, we will also consider the less restrictive condition that the A-terms and Yukawas are simply mutually diagonalizable. This scenario also leads to suppression of FCNC contributions under a rather broad set of conditions.

Giving a precise answer to the question of whether or not the A-terms  
and Yukawas are diagonal in the same basis or, more restrictively,  
whether ${\bf A}_i\sim k_i{\bf Y}_i$ in our setups depends on  
stabilizing the various moduli of our compactification.  However, as  
we will see, we can give an interesting heuristic answer to this  
question with no additional assumptions beyond those we have already  
made. Furthermore, this discussion will point us to other potentially  
interesting constructions. In particular, we will continue to assume  
that the complex structure moduli dependence of the instanton-induced  
operator coefficients can be treated as insignificant $\mathcal{O}(1)$  
factors and that the volume (K\"ahler) moduli can indeed be  
dynamically set to the rough values and hierarchies we take.

Let us focus our discussion on the example with the $U(6)\times SO(2)$  
visible sector and Polonyi hidden sector. Furthermore, we will focus  
on the same region of moduli space as in the discussion above. Namely,  
we will assume a particular hierarchy $r_1\gg r_{2,3}$, so that
\be\label{lim}
1\ll V(\Sigma_{1})\ll V(\Sigma_{2,3})\ll V(\Sigma_1)^2
\ee
This corresponds to a region of moduli space where it costs a  
significant amount of action to go from an instanton with a particular  
set of wrapping numbers to a configuration with one of the wrapping  
numbers increased by one.

Now, let us discuss the Yukawa couplings of the $U(6)\times SO(2)$  
visible sector. Note that the Yukawa couplings of the visible sector  
are of two types. The first type are perturbative couplings from the  
tree level quiver superpotential and are of the form
\be\label{TreeY}
W_{{\rm tree}}=Y_{ijk}A^i\bar{Q}^j\bar{Q}^k
\ee
The second type of terms are non-perturbative Yukawa couplings  
generated by D-instantons and are of the form
\be\label{NPY}
W_{\rm np}\sim e^{-V_1/g_s}(A^1)^3+e^{-V_2/g_s}(A^2)^3+e^{-V_3/g_s} 
(A^3)^3\sim e^{-V_1/g_s}(A^1)^3
\ee
in our approximation. We have used the shorthand $(A^1)^3= 
\epsilon^{a_1b_1a_2b_2a_3b_3}A^1_{a_1b_1}A^1_{a_2b_2}A^1_{a_3b_3}$, etc. As  
mentioned in the example, these non-perturbative Yukawas give rise to  
the u-type quark couplings \cite{Ibanez:2007tu}.

By comparing (\ref{soft_Polonyi}) with the Yukawa couplings we have  
just described above, it should be clear that the A-term and Yukawa  
matrices are highly non-aligned. Indeed, it is not hard to see why  
this is the case. First of all, the mediating  
instantons do not generate A-terms that correspond to the perturbative  
Yukawas since the perturbative superpotential comes from closed paths  
in the quiver, while the instanton-induced terms come from open paths  
(which we define to include two-tensor field loops at the same node).  
Note that this lack of A-terms corresponding to the perturbative  
Yukawa couplings is not a problem since it does not affect the mutual  
diagonalizability of the A-terms and Yukawas. Furthermore, if we want, we can presumably  
generate such terms by going to regions of the moduli space where  
instanton mediation and e.g. gauge or anomaly mediation are comparable  
in strength---these other mediation mechanisms will generate the \lq  
perturbative' A-terms in a flavor blind way.\footnote{We will briefly discuss the possibility of also generating such soft terms via instantons in the next section.}

Thus it remains only to discuss the mutual diagonalizability and  
alignment of the instanton-induced A-terms with the non-perturbative  
Yukawas. Notice that these couplings, like the corresponding A-terms,  
cannot be generated perturbatively since they violate the anomalous  
$U(1)$ factor of the $U(6)$ node, and so they must be generated by a  
non-perturbative effect like D-brane instantons. Examining our above  
results, it should be obvious that though the A-terms and non-perturbative  
Yukawas are not aligned, they are mutually diagonal!\footnote{A similar  
conclusion applies to our first model.}

Let us press on and try to understand the lack of alignment  
between the A-terms and the Yukawas. This goal is useful because a  
better understanding of this lack of alignment will lead us to a  
slightly more interesting characterization of D-brane instantons that  
may serve as a simple guide in building more complicated models.\footnote{Another motivation is that we could presumably have  
considered moving the visible sector in the first example from  
$(1,0,0)$ to $(1,1,1)$ while keeping the hidden sector fixed at the  
origin. If we again assumed $r_1\sim r_2\sim r_3$, then the A-terms  
and Yukawas would not have been mutually diagonal.}

To that end, note that the non-perturbative Yukawas we have written  
above are generated by the instantons wrapping the cycles $z_i=0$ with  
the dominant contribution coming from the instanton wrapping $z_1=0$.  
We can then see a more general geometric reason for the lack of  
alignment in the non-perturbative sector of the theory: since the  
orientation of the instanton picks out the flavor of the fields it  
couples to, in order to have alignment of the A-terms and the Yukawa  
couplings, the orientations of their generating instantons must also  
align. It is rather easy to see this is not possible by the following  
simple argument. Suppose we could choose a Yukawa generating instanton  
to align with an A-term generating instanton. Then, the two instantons  
would share a common normal $T^2$ which we denote $T^2_N$. On $T^2_N$,  
the Yukawa and mediating instanton worldvolumes are localized at  
points $x_Y$ and $x_M\ne x_Y$ respectively. Since the mediating  
instanton intersects both the hidden and visible sectors, both of  
these sectors must also be localized at $x_M$.\footnote{This last  
statement need not apply in cases involving D7 branes.} Hence, the  
Yukawa instanton cannot intersect the visible sector and no term is  
generated. Still, one might hope that it is possible to approximately  
align the instantons  and hence circumvent our previous argument.

The situation seems better when some of  
the cycles are small since then one could hope to find greater  
alignment by considering instantons that differ by wrappings on these small cycles.
This is not the case if one of the two instantons does not wrap the small
cycle. For example, consider $r_1 \gg r_2$ and two instantons, wrapping $z_1=0$ and
$z_1/r_1 + n \, z_2/r_2=0$ (with $n \in \IZ$), respectively. While it is true that not only both cycles are almost aligned 
but also their volumes are very similar, the generated A-terms and Yukawa couplings are very
different. This is because the first instanton generates a term involving only $A^1$, while the 
second one gives rise to a contribution which mostly depends on $A^2$.

\subsection{A broader definition of instanton orientation}

One potentially interesting solution to the lack of alignment between the mediating and Yukawa instantons is to realize that we have been considering particularly simple instantons---those that are invariant under the orientifold and therefore carry an $O(1)$ CP bundle. These instantons are interesting since the orientifold lifts additional neutral fermionic zero modes that would otherwise lead to a vanishing contribution to the A-terms and Yukawa couplings. However, if we are willing to consider, for example, using fluxes to lift the extra neutral zero modes of the $E3$ branes, then we are free to consider instantons with a $U(1)$ CP bundle. In particular, these instantons can have non-trivial CP orientation given by
\be\label{CPor}
\gamma_{\theta, E3}=\alpha^j
\ee
Therefore, such an instanton carries two orientations: the geometrical orientation we have discussed above and the CP orientation just described.\footnote{Strictly speaking, the CP orientation represents the cycle wrapped by the instanton in the twisted homology of the singularity.} We can use this richer structure to align the instantons geometrically by noting that 

\begin{figure}
\begin{center}
\psfrag{X}[cc][][1]{$X_{ij}$}
\psfrag{SUi}[cc][][.9]{$SU(N)_i$}
\psfrag{SUj}[cc][][.9]{$SU(N)_j$}
\psfrag{a}[cc][][1]{$\alpha$}
\psfrag{b}[cc][][1]{$\beta$}
\includegraphics[width=15cm]{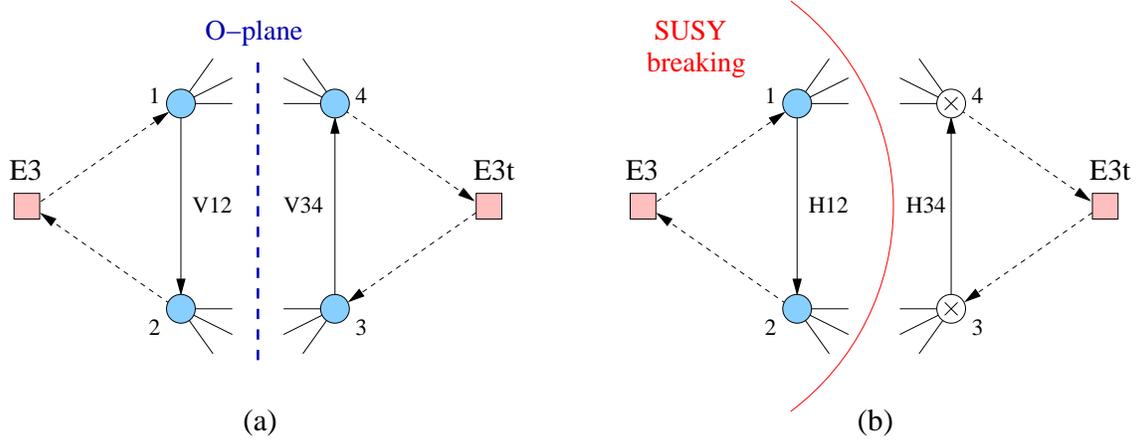}

\caption{The zero mode structure of a $U(1)$ mediating instanton (E3)  
and a Yukawa-generating instanton ($\tilde{\rm E}3$) wrapping the same  
cycle but with different CP orientations. Figure (a) represents the  
interaction structure in the visible sector where the interactions are  
identified by the orientifold plane while Figure (b) represents the  
interaction structure in the hidden sector.}
\label{aligned}
\end{center}
\end{figure}

\begin{itemize}
\item The (untwisted) geometrical orientation controls which flavors the instanton couples to.

\item The CP orientation controls which gauge nodes the instanton interacts with.
\end{itemize}
Hence a simple way to potentially align A-terms and Yukawa couplings is to take their generating instantons to wrap the {\it same} cycle but give the instantons different CP orientation. This means that the instantons will couple to different nodes in both the hidden and visible sector. By considering an orientifolded visible sector, it is possible to identify the different nodes so that the operators the instantons generate are the same in the visible sector. If, however, we take the hidden sector to be non-invariant under the orientifold, then we could imagine the situation in Figure \ref{aligned}, where the mediating instanton couples to fields responsible for SUSY breaking, while the Yukawa instanton couples to empty nodes and hence generates a term without hidden sector fields (note that there are no additional zero modes going between the Yukawa instanton and the hidden sector and so one does not have to worry about a vanishing contribution to the effective action). This strategy may work in higher-order orbifolds and their partial resolutions or in situations where one considers multiple orbifolds of a given space. It would be interesting to find an explicit construction realizing this idea.

\section{Further possibilities}

\label{section_more_ingredients}

We have deliberately kept the previous examples as simple as possible,  only using D3-branes on toroidal orientifolds. There are various refinements that can be introduced in order to obtain more interesting models. We now mention a few of them. It would certainly be interesting to explore model building along these more general lines.

One such extension consists of considering not only orbifold but more general singularities. A practical way of generating many examples of this sort consists of starting from a large orbifold group, for example $\IZ_M \times \IZ_N$, and then partially resolving some of the (orientifold) singularities. We can take this approach to generate more general visible and hidden sectors. Consistency is not affected by partial resolution, since cancellation of twisted and untwisted tadpoles is preserved in the process. The resulting turning on of background values for the twisted K\"ahler moduli might also have some interesting effects on the instanton dynamics.

Another illuminating avenue might be to consider more general compact geometries than $T^6$, since in the case of $T^6$ we have a very simple untwisted cohomology structure. This limited structure can be interesting in the sense that we can then see clear and rather simple connections between completely different physics as in the second example where the need to break SUSY required a certain hierarchy of scales that then was imprinted on the A-terms and the u-type Yukawas.\footnote{In a very limited sense, the conditions required to break SUSY in the hidden sector of this model \lq explain' the relatively large top quark Yukawa coupling!} On the other hand, such a geometry may be unduly restrictive when trying to generate different models of phenomenological interest that avoid troublesome aspects like large FCNCs.

Another possibility is that the compactification might also involve some anti D-brane sectors, which give rise to additional sources of SUSY breaking.

A further extension might be to introduce D7-branes (and anti D7-branes as needed to cancel untwisted tadpoles). D7-branes have various useful applications. For example, they are necessary in simple supersymmetric extensions of the SM based on D-branes at singularities \cite{Aldazabal:2000sa}. They can also give rise to simple metastable SUSY breaking hidden sectors \cite{Franco:2006es,GarciaEtxebarria:2007vh}.

Also, it should be rather simple to construct models that generate B-terms as well. It might then be interesting to study the $\mu/B\mu$ problem in this context.

An direction worth pursuing is to consider more general kinds of E3-brane instantons than those we have considered in the examples. One avenue is to turn on fluxes as a means of both stabilizing the complex structure moduli of the geometry and of lifting the accidental zero modes of the instantons. Also, considering E3-branes with non-trivial gauge bundles would also potentially be interesting, and one could then make contact, via T-duality, with the study of instanton stability and dynamics across lines of marginal stability in the moduli space discussed in \cite{GarciaEtxebarria:2008pi}. A detailed investigation of these topics may lead to a richer set of examples of instanton-generated soft terms and also to a better understanding of instanton-mediation in the closed string picture.

\begin{figure}
\begin{center}
\includegraphics[width=12cm]{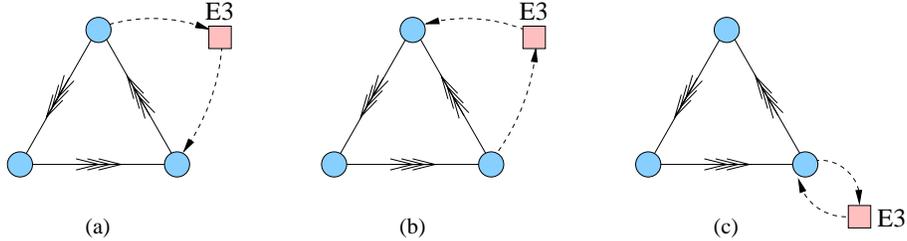}

\caption{The three classes D-brane instantons on $\IC^3/\IZ_3$. Notice the opposite orientation of the fermionic zero modes between (a) and (b). While class (a) couple to single bifundamentals, class (b) couple to linear combinations of products of two of them. Class (c) are completely non-chiral.}
\label{quiver_general_paths}
\end{center}
\end{figure}

We have focused on D-brane instantons wrapping 4-cycles
of the form \eref{E3cycletotal}, which produce operators made out of a 
some linear combination of bifundamental fields connecting a single pair
of nodes in the quiver. In general compactifications, as we mentioned briefly in section 2.1,
we can expect to generate operators of the form \eref{W_inst} (and generalizations for
cases with orientifold identifications), for $X_{ij}$ being an arbitrary oriented path
in the quiver $X_{ij}=X_{i\, k_1}X_{k_1 k_2} \ldots X_{k_n j}$. These more general operators
expand the range of model building possibilities, for example relaxing the conditions for quadratic
and cubic superpotential terms listed in section 2.3. In local constructions, 
i.e. leaving aside the issue of how 4-cycles are completed in a compactification, the question of which
4-cycles are wrapped by the corresponding instantons can be understood in detail. A systematic 
construction of such embeddings for toric singularities can be found in \cite{FGU} (see also appendix A of 
\cite{Franco:2008jc} for a relevant discussion in the related context of flavor D7-branes). \fref{quiver_general_paths} shows the extended quivers 
for general instantons in a local $\IC^3/\IZ_3$ singularity. In this paper, we have considered the first possibility.  

For closed paths in the quiver, i.e. for $i=j$, we generate the determinant of a ``mesonic" operator $X_{ii}$.\footnote{Notice 
that we do not sum over initial and final $SU(N)^{(i)}$ color indices.} These operators are not perturbatively forbidden
by global $U(1)$ symmetries, since they are neutral under all of them. The corresponding instanton contains vector like 
fermionic zero modes $\alpha$ and $\beta$ (see e.g. \fref{quiver_general_paths}.c), whose mass is controlled by $X_{ii}$ 
according the action term \eref{cubic_coupling_instanton}. In other words, $X_{ii}$ measures the distance between the D3 and 
the E3. The couplings in \cite{Baumann:2006th} are examples of such ``mesonic" operators. Our mechanism can be regarded as 
the open string channel interpretation of the closed string  gravitational exchange in \cite{Baumann:2006th}.

\section{Conclusions}
In this paper we have described a new way of generating soft terms in string compactifications. It is quite interesting to note that instanton mediation has aspects of both open and closed string mediation. On the one hand, it is not sensitive to global $U(1)$ symmetries that constrain (low energy effective) open string mediation, but on the other it is sensitive to the chiral gauge invariants of the various sectors in the theory---in particular, certain hidden sector theories seemingly can never communicate their SUSY breaking to the visible sector via instantons since in these cases instantons project onto trivial chiral gauge invariants.\footnote{An example of this statement is the $SU(5)$ model considered in \cite{Lykken:1998ec}. In this case, the simple $O(1)$ instantons we have focused on would project onto gauge invariants that correspond to determinants of $5\times 5$ anti-symmetric matrices that must trivially vanish.}

In any case, we hope to have given a flavor of instanton mediation in this paper, and we leave it to future work to resolve the various outstanding questions we have raised and find more complete realizations of the ideas we have discussed. Above all, though, we simply hope to have illustrated the point that instanton-mediated physics between various D-brane sectors is rather generic in string compactifications and may serve as a phenomenological constraint on string model building.

\begin{center}
\bf{Acknowledgements}
\end{center}
\medskip

We would like to thank J. Maldacena, M. Papucci, D. Shih, A. Uranga, and H. Verlinde for useful discussions. M.B. is supported by an NSF Graduate Research Fellowship and NSF Grant PHY-0756966. S.F. is supported by the DOE
under contract DE-FG02-91ER-40671. S.F. would like to thank the University of Texas at Austin (NSF Grant No. PHY-0455649) and the 
Michigan Center for Theoretical Physics for hospitality during
part of this project.



\end{document}